\documentclass{article}

\usepackage{arxiv}  % https://github.com/kourgeorge/arxiv-style
\usepackage{adjustbox}  % helps with vertical alignment of subfloats
\usepackage{amsmath,amssymb,amsfonts}  % amssymb for \mathcal, amssymb for \triangleq
\usepackage{amsthm}  % proof environment
\usepackage{booktabs}  % for toprule, midrule and other table formatting commands
\usepackage{forest}  % nice tree diagrams within tikz
\usepackage{geometry}
\usepackage{graphicx}
\usepackage[hidelinks,bookmarks]{hyperref}  % hidelinks removes colour and border
\usepackage{mathdots}  % for \iddots
\usepackage{mathtools}  % for \DeclarePairedDelimiter (https://tex.stackexchange.com/a/118217)
\usepackage{multirow}  % multirow cells in tables
\usepackage[defaultlines=2,all]{nowidow}  % prevent widow and orphan lines
\usepackage{pgfplots}
\usepackage{siunitx}
\usepackage[caption=false,font=normalsize,labelfont=sf,textfont=sf]{subfig}  % options as specified by IEEE (note: can't use subcaption with IEEEtran)
\usepackage{capt-of}  % for \captionof
\usepackage{textcomp}  % IEEE template
\usepackage{tikz}
\usepackage{url}  % helpful for formatting URLs (especially around line breaks and within references)
\usepackage{xcolor}
\usepackage[capitalize,nameinlink]{cleveref}  % cref, includuing with theorems (import before algpseudocode to avoid some warnings, but always after hyperref)
\usepackage{algpseudocode}  % algorithmic environment and customisation of line numbers
\usepackage[noadjust]{cite}
\usepackage[textsize=footnotesize,textwidth=10mm]{todonotes}
\usepackage[clock]{ifsym}

\usetikzlibrary{
    shapes.misc,  % star
    matrix,  % matrix of nodes
    decorations.pathreplacing,  % large curly brace
    backgrounds,  % "on background layer"
    patterns,  % https://tex.stackexchange.com/a/54465
}

\usepgfplotslibrary{statistics, fillbetween, groupplots}

\newcommand{\norm}[1]{\left\lVert#1\right\rVert}

\definecolor{colour1}{HTML}{006ed4}  % blue
\definecolor{colour2}{HTML}{e41a1c}  % red
\definecolor{colour3}{HTML}{984ea3}  % purple
\definecolor{colour4}{HTML}{50C878}  % green
\definecolor{colour5}{HTML}{ff7f00}
\definecolor{colour6}{HTML}{a65628}
\definecolor{colour7}{HTML}{f781bf}
\definecolor{colour8}{HTML}{999999}

\definecolor{colourAlgoComment}{HTML}{787878}
\colorlet{lightgray}{black!3}

\tikzset{
	every pin/.style={
		pin edge={black,thick},
		font=\footnotesize
	},
}

\newcommand{\algocomment}[1]{\hfill\textcolor{colourAlgoComment}{// \textit{#1}}}  % standard in-line comment
\algnewcommand{\LineComment}[1]{\textcolor{colourAlgoComment}{// \textit{#1}}}  % custom full line comment

\newcounter{algoCounter}  % custoom counter for algorithms
\algnewcommand{\IfThen}[2]{% \IfThen{<if>}{<then>} ... single-line If/Then
    \State \algorithmicif\ #1\ \algorithmicthen\ #2}

\makeatletter
\newcommand*{\algrule}[1][\algorithmicindent]{%
    \hspace*{.5em}% <------------- This is where the rule starts from
    \vrule %height .75\baselineskip depth .25\baselineskip
    \hspace*{\dimexpr#1-.2em-.4pt}%
}

\newcommand{\StatePar}[1]{%
    \State\parbox[t]{\dimexpr\linewidth-\ALG@thistlm}{\strut #1\strut}%
}
\renewcommand{\ALG@beginalgorithmic}{\offinterlineskip}% Remove all interline skips

\newcount\ALG@printindent@tempcnta
\def\ALG@printindent{%
    \ifnum \theALG@nested > 0% is there anything to print
    \ifx\ALG@text\ALG@x@notext% is this an end group without any text?
    \else
    \unskip
    \ALG@printindent@tempcnta=1
    \loop
    \algrule[\csname ALG@ind@\the\ALG@printindent@tempcnta\endcsname]%
    \advance \ALG@printindent@tempcnta 1
    \ifnum \ALG@printindent@tempcnta<\numexpr\theALG@nested+1\relax
    \repeat
    \fi
    \fi
}
\patchcmd{\ALG@doentity}{\noindent\hskip\ALG@tlm}{\ALG@printindent}{}{\errmessage{failed to patch}}
\makeatother

\algrenewcommand\algorithmicend{\strut\textbf{end}}
\algrenewcommand\algorithmicdo{\strut\textbf{do}}
\algrenewcommand\algorithmicwhile{\strut\textbf{while}}
\algrenewcommand\algorithmicfor{\strut\textbf{for}}
\algrenewcommand\algorithmicforall{\strut\textbf{for all}}
\algrenewcommand\algorithmicloop{\strut\textbf{loop}}
\algrenewcommand\algorithmicrepeat{\strut\textbf{repeat}}
\algrenewcommand\algorithmicuntil{\strut\textbf{until}}
\algrenewcommand\algorithmicprocedure{\strut\textbf{procedure}}
\algrenewcommand\algorithmicfunction{\strut\textbf{function}}
\algrenewcommand\algorithmicif{\strut\textbf{if}}
\algrenewcommand\algorithmicthen{\strut\textbf{then}}
\algrenewcommand\algorithmicelse{\strut\textbf{else}}

\algrenewcommand\algorithmicrequire{\strut\textbf{Input:}}
\algrenewcommand\algorithmicensure{\strut\textbf{Output:}}

\let\oldState\State
\renewcommand{\State}{\oldState\strut}

\pgfplotsset{
	compat=1.17,
	every axis/.style={
        axis on top,
		font=\footnotesize,
		table/col sep=comma,
		tick align=outside,
        xtick = data,
        xtick pos=bottom,
        ytick pos=left,
		log origin=infty,  % bars drown from bottom
        log ticks with fixed point,
        legend cell align={left},
        legend style={
            legend pos=north east,
            align=left,
            row sep=0em,
            column sep=0.2em,
            font=\footnotesize,
            fill opacity=0.75,
            text opacity=1,
            /tikz/nodes={inner sep=0.1em},
            /tikz/every odd column/.style={yshift=0.1em},
        },
		title style={yshift=-0.7em},
        scale only axis,  % width and height apply only to plot area, not labels/ticks/titles
        xticklabel style={text height=0.5em},  % fix vertical alignment of labels
	},
    generic bar plot/.style={
        scaled y ticks = false,
        enlarge x limits=0.25,
        bar width=10pt,
        ybar=\pgflinewidth,
        ymin=0,
        area legend,
        legend image post style={scale=0.6},
        cycle list={
            {white, fill=colour1!70},
            {white, fill=colour2!70},
			{white, fill=colour3!70},
			{white, fill=colour4!70},
			{white, fill=colour5!70},
			{white, fill=colour6!70},
			{white, fill=colour7!70},
			{white, fill=colour8!70}
        },
    },
    generic line plot/.style={
        cycle list={
            {colour1!70, thick, mark options={opacity=1, scale=0.9}, mark=*},
            {colour2!70, thick, mark options={opacity=1, scale=0.8}, mark=square*},
			{colour3!70, thick, mark options={opacity=1, scale=1.1}, mark=triangle*},
			{colour4!70, thick, mark options={opacity=1, scale=1.0}, mark=pentagon*},
			{colour5!70, thick, mark options={opacity=1, scale=1.1}, mark=diamond*},
			{colour6!70, thick, mark options={opacity=1, scale=1.0}, mark=oplus*},
			{colour7!70, thick, mark options={opacity=1, scale=1.0}, mark=10-pointed star},
			{colour8!70, thick, mark options={opacity=1, scale=0.8}, mark=halfsquare*}
        },
    },
    generic scatter plot/.style={
		only marks,
		cycle list={
			{colour1, mark options={line width=0.5pt, fill opacity=0.45, draw opacity=0.7, scale=1.55}, mark=*},
			{colour2, mark options={line width=0.5pt, fill opacity=0.45, draw opacity=0.7, scale=1.33}, mark=square*},
			{colour3, mark options={line width=0.5pt, fill opacity=0.45, draw opacity=0.7, scale=2.0}, mark=triangle*},
			{colour4, mark options={line width=0.5pt, fill opacity=0.45, draw opacity=0.7, scale=1.55}, mark=pentagon*},
			{colour5, mark options={line width=0.5pt, fill opacity=0.45, draw opacity=0.7, scale=1.7}, mark=diamond*},
			{colour6, mark options={line width=0.5pt, fill opacity=0.45, draw opacity=0.7, scale=1.5}, mark=oplus*},
			{colour7, mark options={line width=0.5pt, fill opacity=0.45, draw opacity=0.7, scale=1.5}, mark=10-pointed star},
			{colour8, mark options={line width=0.5pt, fill opacity=0.45, draw opacity=0.7, scale=1.33}, mark=halfsquare*}
		},
    },
    paired bar plot/.style={
        generic bar plot,
        bar width=7pt,
        cycle list={
            {white, fill=colour1!70},
            {white, fill=colour1!70, postaction={pattern=north east lines, pattern color=white}},
            {white, fill=colour2!70},
            {white, fill=colour2!70, postaction={pattern=north east lines, pattern color=white}},
            {white, fill=colour3!70},
            {white, fill=colour3!70, postaction={pattern=north east lines, pattern color=white}},
            {white, fill=colour4!70},
            {white, fill=colour4!70, postaction={pattern=north east lines, pattern color=white}},
            {white, fill=colour5!70},
            {white, fill=colour5!70, postaction={pattern=north east lines, pattern color=white}},
        }
    },
}
\newcommand{\coefficient}{\beta}
\newcommand{\coefficients}{\boldsymbol{\coefficient}}
\newcommand{\baseline}{\coefficient_0}
\newcommand{\sensitivity}{\coefficient_1}
\newcommand{\dist}{d}
\newcommand{\kernelsigma}[1]{\sigma_{f}^{#1}}
\newcommand{\nobs}{N}
\newcommand{\oxygen}{O$_2$}
\newcommand{\Tcalibrations}{\boldsymbol{t}_c}

\newcommand{\tf}[1]{g(#1)}

\newcommand{\calibrationintervalinitial}{T_{0}}
\newcommand{\calibrationintervalmax}{T_{\mathrm{max}}}
\newcommand{\frequency}{f}
\newcommand{\frequencyavailable}{\frequency_{\Delta}}
\newcommand{\frequencyinstant}{\frequency'(\currenttime)}
\newcommand{\frequencymin}{\frequency_{\mathrm{min}}}
\newcommand{\frequencynew}{\frequency(\currenttime)}
\newcommand{\frequencyold}{\frequency(t_{i-1})}
\newcommand{\learningrate}{\alpha}
\newcommand{\ncalibrations}{n}
\newcommand{\ncoefficients}{k}
\newcommand{\nsensors}{M}
\newcommand{\sensors}{\mathcal{S}}
\newcommand{\uncertainty}{u}
\newcommand{\uncertaintynorm}{\uncertainty_{\mathrm{norm}}(\currenttime)}
\newcommand{\uncertaintytotal}{\uncertainty_{\mathrm{total}}(\currenttime)}
\newcommand{\currenttime}{t_i}

\newcommand{\sensorTwoMissingData}{
    \draw[fill=gray!25,draw=none] (axis cs:34.73,-200) rectangle (34.97,1000);
    \draw[fill=gray!25,draw=none] (axis cs:35.00,-200) rectangle (37.50,1000);
    \draw[fill=gray!25,draw=none] (axis cs:50.77,-200) rectangle (50.90,1000);
    \draw[fill=gray!25,draw=none] (axis cs:56.87,-200) rectangle (57.14,1000);
    \draw[fill=gray!25,draw=none] (axis cs:59.40,-200) rectangle (59.74,1000);
    \draw[fill=gray!25,draw=none] (axis cs:59.77,-200) rectangle (61.10,1000);
    \draw[fill=gray!25,draw=none] (axis cs:61.14,-200) rectangle (64.10,1000);
    \draw[fill=gray!25,draw=none] (axis cs:71.87,-200) rectangle (73.00,1000);
    \draw[fill=gray!25,draw=none] (axis cs:73.14,-200) rectangle (73.44,1000);
    \draw[fill=gray!25,draw=none] (axis cs:73.54,-200) rectangle (73.74,1000);
    \draw[fill=gray!25,draw=none] (axis cs:73.77,-200) rectangle (73.90,1000);
    \draw[fill=gray!25,draw=none] (axis cs:190.07,-200) rectangle (190.17,1000);
    \draw[fill=gray!25,draw=none] (axis cs:194.77,-200) rectangle (195.04,1000);
    \draw[fill=gray!25,draw=none] (axis cs:224.74,-200) rectangle (224.84,1000);
    \draw[fill=gray!25,draw=none] (axis cs:355.01,-200) rectangle (355.25,1000);
    \draw[fill=gray!25,draw=none] (axis cs:355.28,-200) rectangle (355.38,1000);
    \draw[fill=gray!25,draw=none] (axis cs:356.95,-200) rectangle (358.41,1000);
    \draw[fill=gray!25,draw=none] (axis cs:366.11,-200) rectangle (371.31,1000);
    \draw[fill=gray!25,draw=none] (axis cs:371.78,-200) rectangle (372.01,1000);
    \draw[fill=gray!25,draw=none] (axis cs:372.05,-200) rectangle (372.21,1000);
    \draw[fill=gray!25,draw=none] (axis cs:372.31,-200) rectangle (373.18,1000);
    \draw[fill=gray!25,draw=none] (axis cs:373.21,-200) rectangle (373.51,1000);
    \draw[fill=gray!25,draw=none] (axis cs:373.88,-200) rectangle (374.18,1000);
    \draw[fill=gray!25,draw=none] (axis cs:379.98,-200) rectangle (380.51,1000);
    \draw[fill=gray!25,draw=none] (axis cs:380.78,-200) rectangle (381.38,1000);
    \draw[fill=gray!25,draw=none] (axis cs:381.45,-200) rectangle (382.18,1000);
    \draw[fill=gray!25,draw=none] (axis cs:382.75,-200) rectangle (382.88,1000);
    \draw[fill=gray!25,draw=none] (axis cs:383.21,-200) rectangle (384.51,1000);
    \draw[fill=gray!25,draw=none] (axis cs:384.81,-200) rectangle (385.05,1000);
    \draw[fill=gray!25,draw=none] (axis cs:386.81,-200) rectangle (386.91,1000);
    \draw[fill=gray!25,draw=none] (axis cs:389.25,-200) rectangle (390.25,1000);
    \draw[fill=gray!25,draw=none] (axis cs:390.45,-200) rectangle (391.05,1000);
    \draw[fill=gray!25,draw=none] (axis cs:391.08,-200) rectangle (392.55,1000);
    \draw[fill=gray!25,draw=none] (axis cs:392.58,-200) rectangle (392.78,1000);
    \draw[fill=gray!25,draw=none] (axis cs:392.81,-200) rectangle (394.48,1000);
}

\begin{document}

    \title{Not all those who drift are lost: Drift correction and calibration scheduling for the IoT}

    \author{%
        Aaron Hurst$^{1}$, Andrey V. Kalinichev$^{2,3}$, Klaus Koren$^{2}$ and Daniel E. Lucani$^{1}$
        \\
        $^1$Department of Electrical and Computer Engineering, Aarhus University, Aarhus, Denmark.\\
        $^2$Department of Biology -- Microbiology, Aarhus University, Aarhus, Denmark.\\
        $^3$Aarhus Institute of Advanced Studies, Aarhus University, Aarhus, Denmark.\\
        \texttt{\{ah, daniel.lucani\}@ece.au.dk} \quad \texttt{akalinichev@aias.au.dk} \quad \texttt{klaus.koren@bio.au.dk}
    }

    \maketitle

    \begin{abstract}
    Sensors provide a vital source of data that link digital systems with the physical world.
    However, as sensors age, the relationship between what they measure and what they output changes.
    This is known as sensor drift and poses a significant challenge that, combined with limited opportunity for re-calibration, can severely limit data quality over time.
    Previous approaches to drift correction typically require large volumes of ground truth data and do not consider measurement or prediction uncertainty.
    In this paper, we propose a probabilistic sensor drift correction method that takes a fundamental approach to modelling the sensor response using Gaussian Process Regression.
    Tested using dissolved oxygen sensors, our method delivers mean squared error (MSE) reductions of up to 90\% and more than 20\% on average.
    We also propose a novel uncertainty-driven calibration schedule optimisation approach that builds on top of drift correction and further reduces MSE by up to 15.7\%.
\end{abstract}

\keywords{drift correction, calibration scheduling, chemical sensors, gaussian process regression}

    \section{Introduction}
\label{sec:introduction}

Modern sensors provide critical interfaces between the physical and digital worlds that enable countless applications. %, spanning from critical healthcare devices to smart home appliances.
As sensors have proliferated exponentially, a significant amount of research has focused on addressing challenges resulting from the huge volume of data being generated, including transmission, storage, energy usage and privacy~\cite{Gulati_2022,Nikoui_2020}.
However, we believe more focus should be placed on the quality of sensor data and, in particular, maintaining sensor calibration accuracy.

Calibration links the signal provided by a sensor to the quantity of an analyte or physical parameter being measured.
For chemical sensors, it is possible to define a \textit{response function} that mathematically describes the relationship between the analyte concentration and the electrical signal produced by the sensor in response to it~\cite{Nalakurthi_2024}.
Calibrating chemical sensors involves taking one or more measurements and comparing to ground truth values to estimate the precise form of the response function.
%Sensor manufacturers often provide a template response function with a number of coefficients that must be calibrated.

Over time, changes in the sensor and its environment invariably result in reduced accuracy as the calibration---i.e., the response function---becomes invalid.
This is known as \textit{drift}~\cite{Nalakurthi_2024}, which can be divided into two types.
First, \textit{environmental drift} occurs when aspects of the environment other than the variable being measured affect the sensor output.
For example, a humidity sensor's output changing due to a temperature increase, despite humidity remaining constant.
%This kind of drift can be viewed as an additional dependence in the response function, e.g. $ \hat{y} = f(x;w) $, where $ w $ represents the additional environmental variables that affect the sensor output.
Second, \textit{sensor drift} occurs when the properties of the sensor itself change such that it responds differently in the same environment.
%In this case, the response function must be adjusted to restore accuracy.
Being both non-linear and unpredictable, sensor drift can quickly undermine data quality and its usability for decision making, especially when drift timeframes are short compared to maintenance windows.

\textit{Drift correction} covers a plethora of methods for detecting, modelling and removing drift from sensor data with the goal of increasing accuracy and reducing the need for (re-)calibration.
%Often, the objective of drift correction is framed as reducing the need for calibration.
Despite a large body of previous work, many approaches are limited due to domain-specific assumptions or require large volumes of ground truth data for training.
Furthermore, despite multiple authors highlighting the variability of calibration uncertainty and optimal calibration frequency~\cite{Nalakurthi_2024,Bresnahan_2021,Samuelsson_2023}, these factors have not been considered until now.

% System overview
\begin{figure}
    \centering
    \input{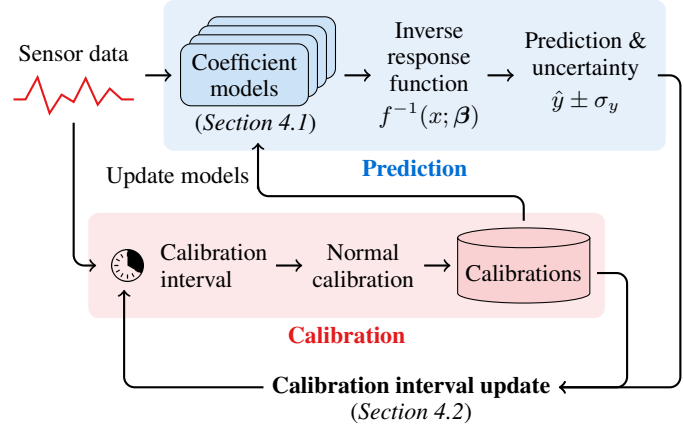}
    \caption{
        Drift correction and calibration interval optimisation framework.
    }
    \label{fig:system}
\end{figure}

In this paper, we make two key contributions.

\textbf{1) Sensor drift correction:}
We propose a novel method for sensor drift correction that, in contrast to previous works, takes a more fundamental approach by directly modelling the evolution of response function coefficients as they drift over time.
More specifically, for each coefficient we build a statistical prediction model using Gaussian Process Regression (GPR) by training on previous calibrations and specifically incorporating calibration uncertainty.
Model predictions are then combined with sensor data via the (inverse) response function to estimate the measured variable and its uncertainty.
In the meantime, calibrations are performed periodically according to sensor-specific calibration intervals.
The proposed approach is illustrated in \cref{fig:system}.

\textbf{2) Calibration Schedule Optimisation:}
We also propose an uncertainty-driven method for optimising calibration scheduling across a collection of sensors that improves accuracy even further.
That is, given a distributed sensor network with heterogeneous drift behaviour and calibration uncertainty, we adaptively optimise sensor-level calibration intervals based on prediction uncertainty to maximise overall accuracy across the network.
In practice, this means that sensors with low uncertainty are calibrated less frequently and high uncertainty sensors are calibrated more often, while holding the total number of calibrations constant.
To the best of our knowledge, no previous works have considered optimising calibration frequency over a sensor network in the context of heterogeneous sensor drift.

Our proposed methods were tested using Clark type electrochemical dissolved oxygen (DO) sensors~\cite{Clark_1953,Revsbech_2021}, which are the most widely used form of DO sensor~\cite{Wei_2019}.
This type of DO sensor measures dissolved oxygen content via the rate at which oxygen is reduced on an electrode surface within the sensor.
DO sensors are critical in many fields, including industrial production, environmental monitoring, food and drug safety and clinical medicine~\cite{Wang_2019a}.
However, they suffer from sensor drift due to issues such as fouling, poisoning and membrane degradation~\cite{Wei_2019,Koren_2023} and have often been overlooked in research.
When applied to a single sensor, our proposed drift correction method reduced mean squared error (MSE) by up to 90\% and by more than 20\% on average.
Network-wide MSE was reduced by a further 11.4\% through calibration interval optimisation.

%Overall, the key contributions of this paper include:
%\begin{enumerate}
%    \item Proposing a broadly applicable method for sensor drift correction that takes a fundamental approach to modelling drift in the sensor response function using GPR and accounts for non-constant calibration uncertainty,
%    \item Proposing a calibration schedule optimisation technique that dynamically adjusts individual sensor calibration intervals based on prediction uncertainty, while maintaining a limited calibration budget, and
%    \item Evaluating our proposed methods on real-world DO sensor data exhibiting highly heterogeneous behaviour.
%\end{enumerate}

The remainder of this paper is structured as follows.
\cref{sec:background} provides necessary background on sensor calibration and GPR.
\cref{sec:relatedwork} discusses related work.
\cref{sec:method} describes our proposed techniques for sensor drift correction.
\cref{sec:evaluation} evaluates of our proposed techniques.
Finally, \cref{sec:conclusion} concludes the paper.
Throughout the paper, vectors are represented using bold lower-case symbols and matrices with non-bold upper-case symbols.
All vectors are column vectors by default.

% Unique/novel parts of our approach (repeated in related work)
%(1) directly modelling drift in the response function,
%(2) incorporating calibration uncertainty,
%(3) enabling model training on only a small number of calibrations, and
%(4) adaptively optimising calibration frequency across a sensor network.

    \section{Background}
\label{sec:background}

%In the following subsections we describe the features of sensor calibration and GPR that are relevant to our work.

\subsection{Sensor Calibration}
\label{sec:background:sensors}

Chemical sensors produce an electrical signal in response to the presence of an analyte.
The relationship between the electrical signal and analyte concentration is characterised by a response function $ \tf{\cdot} $ with coefficients $ \coefficients $ such that
\begin{equation}
    x = \tf{y; \coefficients} + \epsilon,
\end{equation}
where $ x $ is the electrical signal, $ y $ is the analyte concentration and $ \epsilon $ represents noise.
The response function may take any form, but, in a well-designed sensor, should be as simple as possible and ideally linear.

Calibration is the process of taking one or more sensor measurements under conditions where the true analyte concentration is known and estimating the coefficients, $ \coefficients $, from these data, e.g. via regression.
Ground truth values can be obtained by physically controlling analyte concentration or by measuring it using another sensor of known accuracy.

We focus on electrochemical DO sensors, which have an affine response in the majority of their dynamic range, which can be represented as:
\begin{equation} \label{eq:linear_tf}
    \tf{y; \coefficients} = \coefficient_0 + \coefficient_1 y,
\end{equation}
where $ y $ is the concentration or partial pressure of dissolved oxygen (typically expressed in percentage air saturation or \unit{hPa}) and the coefficients $ \coefficient_0 $ and $ \coefficient_1 $ are often referred to as the \textit{baseline} and \textit{sensitivity}, respectively, with units of \unit{mV} and \unit{mV}/\%\oxygen{}.
The raw sensor output $ x $ is typically in picoamperes (\unit{pA}) but often converted to millivolts (\unit{mV}) for read-out and processing.
Two-point calibration is generally sufficient and the coefficients and their uncertainties can be estimated using simple linear regression.
Sensor drift corresponds to changes in the coefficient values over time.

\subsection{Gaussian Process Regression}
\label{sec:background:gpr}

A Gaussian Process is simply a collection of jointly Gaussian random variables~\cite{Rasmusen_2006}.
In the context of regression, i.e., GPR, each random variable represents the predicted output $ y $ at a given value $ \boldsymbol{x} $ in the input domain~\cite{Wang_2023b}.
To provide a prediction over a continuous input domain, the number of random variables must therefore be infinite.
As such, GPR is non-parametric.
Realisations of the random variables correspond to \textit{functions} over the input space.
GPR is therefore often viewed as a distribution over functions~\cite{Rasmusen_2006,Schulz_2018} and, in this sense, is fully specified by a mean $ m(\boldsymbol{x}) $ and covariance function $ k(\boldsymbol{x}, \boldsymbol{x'}) $, often written as follows:
\begin{equation}
    f(\boldsymbol{x}) \sim \mathcal{GP}\left( m(\boldsymbol{x}), k\left(\boldsymbol{x}, \boldsymbol{x'}\right) \right).
\end{equation}

The covariance function, or \textit{kernel}, encodes assumptions about the underlying behaviour of the data being analysed.
Common kernels include the Radial Basis Function (RBF), rational quadratic (RQ) and Matérn kernels, which are defined as follows:
\begin{align}
    k_{\mathrm{RBF}}(\boldsymbol{x},\boldsymbol{x'})
    &= \kernelsigma{2} \exp{\left(- \frac{\dist{}^2}{2 \ell^2}\right)},
    \\
    k_{\mathrm{RQ}}(\boldsymbol{x},\boldsymbol{x'})
    &= \kernelsigma{2} \left( 1 + \frac{\dist{}}{2 \alpha \ell^2} \right)^{-\alpha}, \text{ and}
    \\
    k_{\mathrm{M}}(\boldsymbol{x},\boldsymbol{x'})
    &= \kernelsigma{2} \frac{2^{1-\nu}}{\Gamma(\nu)} \left( \sqrt{2\nu} \frac{\dist{}}{\ell} \right)^{\nu} K_{\nu} \left( \sqrt{2\nu} \frac{\dist{}}{\ell} \right),
\end{align}
where $ \dist{} $ is the Euclidean distance $ \norm{\boldsymbol{x} - \boldsymbol{x'}} $ and $ \kernelsigma{} $, $ \ell $, $ \alpha $ and $ \nu $ are hyper-parameters.
In particular, $ \kernelsigma{} $ controls the kernel variance and $ \ell $ is known as the \textit{characteristic length} and controls the horizontal extent of the kernel and thereby the smoothness of the fitted function.
%controlling the vertical and horizontal extent of the kernel, respectively.
%That is, $ \kernelsigma{} $ controls the variance of the distribution and $ \ell $, also known as the \textit{characteristic length}, controls how smooth the functions are. % the minimum interval over which significant changes in the functions can occur

GPR is a Bayesian approach.
Hence, performing regression begins with defining the prior distribution over observed and predicted outputs.
Formally, given a set of $ \nobs{} $ noisy observations $ D = \{ X, \boldsymbol{y}\} = \{ \boldsymbol{x}_i, y_i \}_{i=1}^\nobs{} $,
%where $ y = \tf{\boldsymbol{x}} + \epsilon $ and $ \epsilon $ is additive independent Gaussian noise,
the joint distribution between the observed outputs $ \boldsymbol{y} $ and predicted outputs $ \boldsymbol{y}_{*} $ at $ \nobs_{*} $ unobserved test inputs $ \boldsymbol{X}_* $ according to the prior is~\cite{MacKay_1998,Rasmusen_2006}:
\begin{equation}
    \begin{bmatrix}
        \boldsymbol{y} \\
        \boldsymbol{y}_{*}
    \end{bmatrix}
    \sim \mathcal{N}\left(
    \begin{bmatrix}
        m(X) \\
        m(X_*)
    \end{bmatrix},
    \begin{bmatrix}
        K + \boldsymbol{\sigma}_y^{\top}I & K_* \\
        K_*^{\top} & K_{**}
    \end{bmatrix}
    \right),
\end{equation}
where $ m $ is the mean function; $ K $ is the symmetric $ \nobs{} \times \nobs{} $ covariance matrix constructed by evaluating $ k(\boldsymbol{x}, \boldsymbol{x}) $ for each combination of inputs; $ K_* $ is the $ \nobs{} \times n_* $ matrix constructed by evaluating $ k(\boldsymbol{x}, \boldsymbol{x}_*) $ for each combination of training and test input; $ K_{**} $ is the $ n_* \times n_* $ matrix constructed by evaluating $ k(\boldsymbol{x}_x, \boldsymbol{x}_*) $; $ \boldsymbol{\sigma}_y = \{ \sigma_i^2 \}_{i=1}^{\nobs{}} $ contains the noise variances for each observation; and $ I $ is the $ \nobs{} \times \nobs{} $ identity matrix.
%Note that if noise variance is constant, i.e.,  = sigma for all i, then this canbe simplified to

Deriving the posterior, or \textit{predictive distribution}, for GPR given the above prior is achieved using standard results~\cite{Rasmusen_2006} by conditioning the prior on the observations $ \boldsymbol{y} $, giving
\begin{equation} \label{eq:predictive_distribution}
    \boldsymbol{y}_{*} | \, \boldsymbol{y}, X, X_*
    \sim \mathcal{N}\left( \boldsymbol{\bar{y}}_{*}, \mathrm{cov}\left(\boldsymbol{y}_{*}\right) \right)
\end{equation}
with:
\begin{align} \label{eq:mean_function}
    \boldsymbol{\bar{y}}_{*}
    &= K_*^{\top} \left( K + \boldsymbol{\sigma}_y^{\top} I \right)^{-1} \boldsymbol{y} \text{, and} \\ \label{eq:covariance_function}
    \mathrm{cov}(\boldsymbol{y}_{*})
    &= K_{**} - K_*^{\top} \left( K + \boldsymbol{\sigma}_y^{\top} I \right)^{-1} K_*.
\end{align}
This is the distribution over functions provided by GPR where $ \boldsymbol{\bar{y}}_{*} $ is the mean function, which is used for prediction.
The hyper-parameters $ \kernelsigma{} $ and $ \ell $ are set by maximising the log marginal likelihood.
%, as follows:
%\begin{equation}
%    \kernelsigma{}, \ell = \argmax_{\kernelsigma{}, \ell } \log \left(p(\boldsymbol{y} | X, \kernelsigma{}, \ell)\right)
%\end{equation}
%where
%\begin{equation}
%    content
%\end{equation}
%and p = ... since

%The mean function can be seen as a linear combination of kernel functions, each one centred on a training point.
%\begin{equation}
%    \bar{f}(\boldsymbol{x}_*) = \sum_{i=1}^{\nobs{}} \alpha_i k(x,x_*)
%\end{equation}
%where $ \alpha_i = (K + \sigma_n^2 I)^{-1} y_i $

%The kernel, or covariance function, plays a pivotal role in this smoothing process, encapsulating our prior knowledge about the functions we aim to model
%
%GP is a collection of random variables. How is it smooth, then?
%Covariance function, i.e., kernel.
%
%the mean function derived from the posterior distribution of possible functions is the function used for regression predictions

    \section{Related Work}
\label{sec:relatedwork}

As defined in \cref{sec:introduction}, drift is a sensor malfunction where the sensor response changes independently of the measured variable, resulting in reduced accuracy~\cite{Ren_2024}.
Environmental drift (also known as second-order, measurement or external drift) occurs when sensor output is affected by environmental conditions other than the measured variable~\cite{Geng_2015}; whereas sensor drift (first-order, true or internal drift) refers to sensor output variation caused by physical changes such as ageing~\cite{Ren_2024,Concas_2021}.
Ultimately, calibration is the only cure for drift, however, drift correction reduces the need for calibration by estimating and removing drift from the sensor output~\cite{Concas_2021}.

Environmental drift correction has been proposed for several different domains.
For example, navigational gyroscope systems subject to temperature-related drift can be corrected using Support Vector Regression (SVR)~\cite{Zhao_2022,Zhao_2022a}.
Among many proposed environmental drift correction methods for air pollution, SVR also appears to be most effective~\cite{Aula_2022}.
%Authors~\cite{Aula_2022} benchmarked several techniques for environmental drift correction in air pollution and also found SVR to be most effective.
GPR has been used for simulated chemiresistors with its uncertainty estimation also exploited to optimise calibration~\cite{Geng_2015}
Hardware-based environmental drift correction has been proposed for gas sensors~\cite{Ren_2024} and industrial displacement sensors~\cite{Wang_2022a}.

Sensor drift has likewise been investigated extensively with many techniques proposed.
Neural networks (NN) have been tested in simulations for water quality sensors~\cite{Khatri_2020} and extensively studied for detecting sensor drift in temperature sensors embedded within a footbridge~\cite{Pereira_2023}.
In~\cite{Pereira_2023}, the authors use a probabilistic NN to estimate the true bridge temperature based on external environmental conditions and compare this to the sensor output to detect drift.

One domain with disproportionate focus on sensor drift is gas sensors or so-called electronic noses~\cite{Ren_2024,Rudnitskaya_2018}.
E-noses contain multiple individual sensors, which are aggregated to classify and estimate the concentration of specific gasses.
Standard methods for e-nose sensor drift correction include component correction~\cite{Artursson_2000,Padilla_2010,Ziyatdinov_2010}, domain adaption~\cite{Lu_2022,Yi_2023,Ma_2018,Liu_2014} and machine learning~\cite{Liang_2021a,Ma_2018,Liu_2014}, in particular classifier ensembles~\cite{Vergara_2012,Das_2020}.
In~\cite{Vergara_2012}, the authors use an ensemble of Support Vector Machine (SVM) classifiers trained on different historical data to correct sensor drift.
GPR-based methods have also been proposed, but appear to very slightly underperform SVMs for gas concentration prediction~\cite{Monroy_2012}.
However, GPR offers the unique benefit of incorporating both data and prediction uncertainty.

Spatial interpolation between sensors, on the other hand, has proven to be a fruitful application for GPR~\cite{Anand_2021,Kumar_2013,Persic_2021}.
If reference (ground truth) sensors are present, spatial interpolation can also be used for drift correction~\cite{Zheng_2019}.
Other related applications of GPR include hysteresis correction~\cite{Urban_2015} and sensor health monitoring~\cite{Lee_2019a}, where the use pairwise comparisons between sensors can detect poorly performing sensors.

In summary, while a wide range of drift compensation techniques have been developed, many rely on domain-specific assumptions or large volumes of labelled data.
Our proposed approach, which focuses on sensor drift, addresses these limitations by directly modelling drift in the sensor response function coefficients, making it inherently sensor- and application-agnostic.
To the best of our knowledge, it is the first approach to combine several novel features, including:
(1) explicitly incorporating calibration and prediction uncertainty,
(2) high performance using only a small number of reference measurements (i.e., calibrations), and
(3) adaptive calibration scheduling based on prediction uncertainty.
This stands in contrast to existing methods, particularly other supervised learning approaches, which often require extensive historical datasets for training.
By leveraging the probabilistic nature of GPR, our method provides a principled and data-efficient framework for real-time sensor drift correction in diverse contexts.

    \section{Method}
\label{sec:method}

In this paper, we propose both a novel method for sensor drift correction using GPR and a method for adaptively optimising the sensor-level calibration frequency across a network of sensors based on prediction uncertainty.
These approaches are described in the following subsections.

%In this paper, we approach the challenge of sensor drift from two angles.
%First, we consider the case of a single sensor for which we must correct sensor drift in order to improve estimation accuracy of the measured parameter.
%Second, we consider a scenario with multiple sensors experiencing heterogeneous sensor drift and attempt to optimise the timing and allocation of calibrations between them given a limited budget for performing calibrations.

\subsection{Sensor Drift Correction}
\label{sec:method:drift_correction}

\textbf{Problem 1: Sensor Drift Correction.}
\textit{Given a series of $ \nobs{} $ sensor observations $ \boldsymbol{x} = \{x_i\}_{i=1}^\nobs{} $ at times $ \boldsymbol{t} = \{t_i\}_{i=1}^\nobs{} $ and $ \ncalibrations{} $ calibrations during this period at times $ \Tcalibrations{} = \{t_j\}_{j=1}^{\ncalibrations{}} $, each of which estimates the sensor response function coefficients $ \coefficients(t) = \{\coefficient_0(t), \coefficient_1(t), \ldots\} $ and their uncertainties, estimate the analyte concentration $ y $ at all times in $ \boldsymbol{t} $.}

% estimate how the sensor properties drift over time in order to maximise the accuracy of the drift-corrected measurements $ \boldsymbol{\hat{y}} $.}

We assume that all response function coefficients $ \{\coefficient_0, \coefficient_1, \ldots\} $ vary smoothly and propose using GPR to model their evolution over time using the calibration measurements and uncertainties as training data.
That is, as illustrated in \cref{fig:system}, we separately train a single GPR model for each coefficient where the independent variable is time and the dependent variable is the value of the coefficient.
By making the appropriate substitutions in \cref{eq:predictive_distribution,eq:mean_function,eq:covariance_function}, the predictive distribution for a given coefficient $ \coefficient_i $ at test points $ \boldsymbol{t}_* $ according to GPR is
\begin{equation}
    \boldsymbol{\coefficient{}}_{i*} \, | \, \hat{\boldsymbol{\coefficient{}}_{i}}, \Tcalibrations{}, \boldsymbol{t}_*
    \sim \mathcal{N}\left( \boldsymbol{\bar{\coefficient}}_{i*}, \mathrm{cov}\left(\boldsymbol{\coefficient{}}_{i*}\right) \right)
\end{equation}
with:
\begin{align}
    \boldsymbol{\bar{\coefficient}}_{i*}
    &= K_*^{\top} \left( K + \boldsymbol{\sigma}_{i}^{\top} I \right)^{-1} \hat{\boldsymbol{\coefficient{}}_{i}} \text{, and} \\
    \mathrm{cov}(\boldsymbol{\coefficient{}}_{i*})
    &= K_{**} - K_*^{\top} \left( K + \boldsymbol{\sigma}_{i}^{\top} I \right)^{-1} K_*,
\end{align}
where $ \hat{\boldsymbol{\coefficient{}}_{i}} $ contains all calibrations for coefficient $ \coefficient_i $, $ \boldsymbol{\sigma}_{i} $ contains the uncertainties associated with the calibrations and the matrices $ K $, $ K_* $ and $ K_{**} $ are constructed with the kernel function evaluated over the training data (calibration times) and test points.
We can then model the coefficients as continuous functions of time such that $ \coefficient{}_{i}(t) = \boldsymbol{\bar{\coefficient}}_{i*} $.

The uncertainty, $ \boldsymbol{\sigma}_{i} $, associated with calibration can be estimated in various ways, depending on the particular application.
For our use case (DO sensors with affine response function), calibration is performed using linear regression, so the regression coefficient standard errors can be used to estimate $ \boldsymbol{\sigma}_{i} $.

Prediction of the analyte concentration is achieved by inverting the response function and using the GPR mean functions for the coefficients as follows:
\begin{equation}
    \hat{y}_t = f^{-1}(x_t; \coefficients(t)).
\end{equation}
In the case of an affine response function, this becomes:
\begin{equation}
    \hat{y}_t = \frac{x_t - \coefficient_0(t)}{\coefficient_1(t)}.
\end{equation}

\begin{figure}
    \centering
    \begin{tikzpicture}
    \begin{groupplot}[
        group style={rows=5,xlabels at=edge bottom,vertical sep=0.25cm,xticklabels at=edge bottom},
        generic line plot,
        height=1.2cm,
        width=0.87\columnwidth,
        xtick={0,100,200,300,400},
        xlabel={Time since start of observations, $ t $ (hours)},
        xmin=-10,xmax=416,
        ]
        %%%%%%%%%%%%%%%%%%%%%%%%%%%%%%%%%%%%%%%%%%%%%%%%%%%%%%%%%%%%%%%%%%%%%%%%
        % Data
        \nextgroupplot[ymax=800]
            \addplot [thick,colour1] table[x index=0, y index=2] {data__sensor_data_sparser.csv};
            \sensorTwoMissingData
            \node[anchor=north west,font=\footnotesize] () at (axis cs:75,800) {(a) Sensor data, \unit{mV}};

        %%%%%%%%%%%%%%%%%%%%%%%%%%%%%%%%%%%%%%%%%%%%%%%%%%%%%%%%%%%%%%%%%%%%%%%%
        % Baseline/intercept
        \nextgroupplot
            % Confidence interval
            \addplot [white, name path=L] table[x index=0, y index=3] {data__sensor_2_10_hrs_matern_length_200_data.csv};
            \addplot [white, name path=H] table[x index=0, y index=4] {data__sensor_2_10_hrs_matern_length_200_data.csv};
            \addplot [colour1,fill opacity=0.1] fill between[of=L and H];

            % Prediction
            \addplot [thick,colour1] table[x index=0, y index=2] {data__sensor_2_10_hrs_matern_length_200_data.csv};

            % Calibrations
            \addplot [thick,black,only marks,mark size=1.5pt,mark=*] table[x index=0, y index=1] {data__sensor_2_10_hrs_matern_length_200_calibrations.csv};

            \node[anchor=north west,font=\footnotesize] () at (axis cs:75,45) {(b) Intercept coefficient $ \baseline $, \unit{mV}};

        %%%%%%%%%%%%%%%%%%%%%%%%%%%%%%%%%%%%%%%%%%%%%%%%%%%%%%%%%%%%%%%%%%%%%%%%
        % Sensitivity/slope
        \nextgroupplot[ymin=-2]
            % Confidence interval
            \addplot [white, name path=L] table[x index=0, y index=6] {data__sensor_2_10_hrs_matern_length_200_data.csv};
            \addplot [white, name path=H] table[x index=0, y index=7] {data__sensor_2_10_hrs_matern_length_200_data.csv};
            \addplot [colour1,fill opacity=0.1] fill between[of=L and H];

            % Prediction
            \addplot [thick,colour1] table[x index=0, y index=5] {data__sensor_2_10_hrs_matern_length_200_data.csv};

            % Calibrations
            \addplot [thick,black,only marks,mark size=1.5pt,mark=*] table[x index=0, y index=2] {data__sensor_2_10_hrs_matern_length_200_calibrations.csv};

            \node[anchor=north east,font=\footnotesize] () at (axis cs:416,6.5) {(c) Slope coefficient $ \sensitivity $, \unit{mV}/\%\oxygen{}};

        %%%%%%%%%%%%%%%%%%%%%%%%%%%%%%%%%%%%%%%%%%%%%%%%%%%%%%%%%%%%%%%%%%%%%%%%
        % Drift corrected
        \nextgroupplot[ymax=250]
            \addplot [thick,colour1] table[x index=0, y index=8] {data__sensor_2_10_hrs_matern_length_200_data.csv};
            \sensorTwoMissingData
            \node[anchor=north west,font=\footnotesize] () at (axis cs:75,250) {(d) Drift-corrected, \%\oxygen};

        %%%%%%%%%%%%%%%%%%%%%%%%%%%%%%%%%%%%%%%%%%%%%%%%%%%%%%%%%%%%%%%%%%%%%%%%
        % Residuals
        \nextgroupplot[ymax=70,ymin=-70]
            \addplot [thick,colour1,only marks,mark size=0.2pt,mark=*] table[x index=0, y index=1] {data__sensor_2_10_hrs_matern_length_200_residuals.csv};
            \draw[fill=gray!25,draw=none] (axis cs:74.83,-200) rectangle (238.56,200);
            \node[anchor=center,font=\footnotesize] () at (axis cs:156,0) {No reference data};
            \node[anchor=south west,font=\footnotesize] () at (axis cs:240,-70) {(e) Residuals, \%\oxygen};

    \end{groupplot}
\end{tikzpicture}
    \caption{Drift correction example: (a)~original data, (b,c)~calibrations and GPR models for the response function coefficients, (e)~corrected data and (e)~residuals.}
    \label{fig:example}
\end{figure}
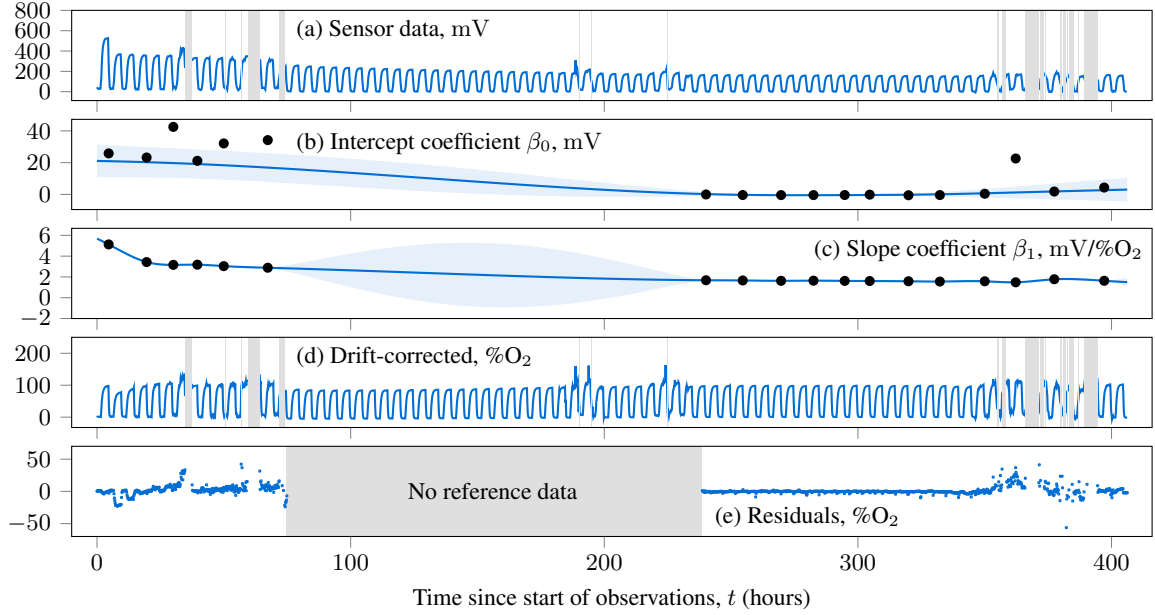

\cref{fig:example} shows an example of our method in action.
In (a), we have raw sensor output, which exhibits significant noise, outliers, invalid data (grey regions) and gradually falling amplitude, indicative of sensor drift.
In (b) and (c), the individual calibrations of the response function coefficients are shown (black dots) along with the fitted GPR models (blue lines) and confidence intervals (blue shaded regions).
Some calibrations, i.e., those that lie outside the model's confidence interval, have less influence over the fitted model, which can be due to their outlying values or high uncertainty.
\cref{fig:example}(d) shows the resulting drift-corrected predictions, which still contain noise, but maintain a steady amplitude range between 0 and 100, as expected according to the ground truth data.
Finally, (e) shows the residuals, of which 50\% are less than $ \pm $1 and 90\% are less than $ \pm $10.
Note that the reference sensor (ground truth) was unavailable during a portion of the experiment, hence the gap in the residuals.

%%%%%%%%%%%%%%%%%%%%%%%%%%%%%%%%%%%%%%%%%%%%%%%%%%%%%%
% Drift correction example plot
%%%%%%%%%%%%%%%%%%%%%%%%%%%%%%%%%%%%%%%%%%%%%%%%%%%%%%

We evaluate the accuracy of drift correction using the mean squared error (MSE) compared to known ground truth values obtained from a separate reference sensor.
That is,
\begin{equation}
    \mathrm{MSE}
    = \frac{1}{\nobs{}} \norm{\boldsymbol{\hat{y} - \boldsymbol{y}}}^2
    = \frac{1}{\nobs{}} \sum_{i=1}^{\nobs{}} \left(\hat{y}_i - y_i\right)^2,
\end{equation}
where $ \hat{y} $ and $ y $ represent the drift-corrected and ground truth values and $ \nobs $ is the number of observations.
Root mean squared error (RMSE) equal to $ \sqrt{MSE} $ is also used for some evaluations.

\textit{Offline vs. Online:}
We consider applying drift correction in both offline and online scenarios.
By offline, we mean that all data are available in advance for training the models.
In contrast, an online scenario requires predictions to be made immediately by relying only on past and current information.
Thus, offline predictions at time $ t $ have the benefit of knowing about future calibrations, whereas online predictions do not.
An offline scenario is relevant when correcting drift in an existing dataset or when data timeliness is less important, whereas online drift correction is relevant for real time predictions.
Naturally, online drift correction is more challenging, especially as the time interval between calibrations increases.

\subsection{Calibration Schedule Optimisation}
\label{sec:method:calibration_interval_optimisation}

\textbf{Problem 2: Calibration Schedule Optimisation.}
\textit{Given a collection of sensors and a limited budget for performing calibrations, optimise the allocation and timing of calibrations across all sensors such that collection-wide measurement accuracy is maximised.}

%%%%%%%%%%%%%%%%%%%%%%%%%%%%%%%%%%%%%%%%%%%
% Calibration frequency update algorithm
\begin{figure}[!t]
    \centering
    \footnotesize
    \begin{tabular}{p{0.94\columnwidth}}
	\toprule
	\textbf{Algorithm 1} Calibration interval update \\ \midrule
	\textbf{Inputs:} current time $ \currenttime $, sensors $ \sensors $, learning rate $ \learningrate $, initial calibration interval $ \calibrationintervalinitial $, maximum calibration interval $ \calibrationintervalmax $ \\
	\vspace{-0.65em}
	\begin{algorithmic}[1]%\setstretch{\algolinespacing}
        \State $ \nsensors \leftarrow \mathrm{sizeof}(\sensors) $  \algocomment{number of sensors}
        \State $ \frequencymin \leftarrow 1 \,/\, \calibrationintervalmax $  \algocomment{minimum calibration frequency}
        \State $ \frequencyavailable \leftarrow \nsensors \left( 1  \,/\, \calibrationintervalinitial - \frequencymin \right) $  \algocomment{disposable frequency}
        \State
        \State \LineComment{Compute total uncertainty across all sensors}
        \State $ \uncertaintytotal \leftarrow 0 $
        \For{sensor in $ \sensors $}
            \State $ \uncertaintytotal \leftarrow \uncertaintytotal $ + sensor.getUncertainty($ \currenttime $)
        \EndFor
        \State
        \State \LineComment{Update sensor calibration intervals}
        \For{sensor in $ \sensors $}
            \State $ \frequencyold \leftarrow 1 \,/ $ sensor.getCalibrationInterval()
            \State $ \uncertaintynorm \leftarrow $ sensor.getUncertainty($ \currenttime $) $ /\, \uncertaintytotal $
            \State $ \frequencyinstant \leftarrow \frequencymin + \frequencyavailable \uncertaintynorm $  \algocomment{instantaneous frequency}
            \State $ \frequencynew \leftarrow \learningrate \frequencyinstant + (1 - \learningrate) \frequencyold $  \algocomment{updated frequency}
            \State sensor.setCalibrationInterval($ 1 \,/\, \frequencynew $)
        \EndFor
	\end{algorithmic}
	\\[-1em] \bottomrule
\end{tabular}

\end{figure}
%%%%%%%%%%%%%%%%%%%%%%%%%%%%%%%%%%%%%%%%%%%

We assume that drift and uncertainty affect the sensors non-uniformly.
That is, some sensors experience more drift than others or experience drift during different time periods.
Our proposed method is based on the simple observation that sensors experiencing less drift generally require fewer calibrations for effective drift correction compared to sensors experiencing more drift.
We therefore propose adjusting each sensor's calibration interval so that it is inversely proportional to its current prediction uncertainty, which is estimated from the variance of the GPR models used for drift correction.

The method is outlined in Algorithm 1.
At a high level, we estimating the current prediction uncertainty for each sensor, compute the total uncertainty across all sensors (lines 6--9) and update each sensor's calibration interval in proportion to its contribution to the total uncertainty (lines 12--18).

%% Inputs
Five inputs are required.
First, the current time, $ \currenttime $.
Second, the array of sensors, $ \sensors $.
Third, the learning rate, $ 0 \le \learningrate \le 1$, which controls how quickly sensor calibration intervals can adapt such that a value of 0 results in fixed intervals with no updates, a value of 1 results in intervals based only on instantaneous uncertainty and intermediate values balance current and past information to determine the calibration interval.
Fourth, the initial calibration interval, $ \calibrationintervalinitial $, which defines the overall calibration budget.
That is, the total calibration frequency across all sensors never exceeds the rate that would occur if every sensor were calibrated according to this initial interval.
Fifth, the maximum calibration interval, $ \calibrationintervalmax $, which limits the maximum time between calibrations for an individual sensor, ensuring that stable sensors are not ignored.

%% Upfront calculations
The algorithm starts by counting the number of sensors, $ \nsensors $, and computing the \textit{minimum frequency}, $ \frequencymin $, and \textit{disposable frequency}, $ \frequencyavailable $.
The minimum frequency is simply the inverse of the maximum calibration interval ($ \calibrationintervalmax $), while the disposable frequency represents the total frequency that can be freely allocated between the sensors, subject to the criteria of fixed calibration budget and the maximum calibration interval.
It is defined as the difference between the initial calibration frequency (i.e., $ 1/\calibrationintervalinitial $) and the minimum calibration frequency, multiplied by the number of sensors (line 3).

%% Total prediction uncertainty
Next, the total prediction uncertainty is computed (lines 6--9).
Prediction uncertainty is estimated for a given sensor at time $ \currenttime $ as the quadratic mean of the relative errors of its response function coefficients divided by the square of total number of calibrations up to time $ \currenttime $.
That is,
\begin{equation} \label{eq:uncertainty}
    \uncertainty(\currenttime)
    =
    \frac{1}{\ncalibrations^2}
    \sqrt{
        \frac{1}{\ncoefficients}
        \sum_{i=0}^{\ncoefficients-1} \left( \frac{\sigma_{i}(\currenttime)}{ \coefficient_{i}(\currenttime) } \right)^2
    }
\end{equation}
where $ \ncalibrations $ is the total number of calibrations for the given sensor, $ \ncoefficients $ is the number of coefficients in the sensor response function,
and $ \coefficient_{i}(\currenttime) $ and $ \sigma_{i}(\currenttime) $ are the predicted value and variance for the coefficient $ \coefficient_{i} $ at time $ \currenttime $ according to the relevant GPR model.
%$ \coefficient_{i}(\currenttime) $ is the predicted value of coefficient $ \coefficient_{i} $ at time $ \currenttime $ according to the GPR model for this coefficient and $ \sigma_{i}(\currenttime) $ is the prediction variance for that coefficient according to the same GPR model.
Division by the square of the total number of calibrations prevents over-prioritisation of high-uncertainty sensors.
This calculation is done within the \texttt{getUncertainty} method, which is used on lines 8 and 14.

%% Calibration interval update
Finally, the calibration intervals are updated for each sensor (lines 12--18).
The previous calibration frequency, $ \frequencyold $, is first calculated as the inverse of the current calibration interval (line 13).
Next, the normalised sensor uncertainty, $ \uncertaintynorm $, is computed, which is equal to the individual sensor's contribution to total uncertainty (line 14).
%That is, the uncertainty obtained using \cref{eq:uncertainty} divided by the sum of uncertainty over all sensors.
Next, the \textit{instantaneous calibration frequency}, $ \frequencyinstant $, is calculated by allocating the disposable frequency proportionally among the sensors according to their normalised uncertainty (line 15).
The new calibration frequency, $ \frequencynew $, is then equal to the exponentially weighted moving average of the instantaneous frequency and the previous calibration frequency (line 16).
Finally, the new calibration interval is applied to the sensor (line 17).

Calibration is triggered for a given sensor when the time elapsed since its most recent calibration exceeds its current calibration interval.
The calibration interval for all sensors is initially set to $ \calibrationintervalinitial $ and then updated periodically using Algorithm~1.
The update frequency is set by the user and, together with $ \learningrate $, controls how quickly sensor calibration intervals adapt to changes in relative uncertainty.
We therefore suggest fixing the update frequency and varying only $ \learningrate $ during hyperparameter tuning.

%%%%%%%%%%%%%%%%%%%%%%%%%%%%%%%%%%%%%%%%%%%%%%%%%%%%%%
% Drift correction example plot
\begin{figure}
    \centering
    \begin{tikzpicture}
    \begin{axis}[
        generic line plot,
        no marks,
        height=2.0cm,
        width=0.82\columnwidth,
        xtick={0,100,200,300,400},
        ytick={0,50,100},
        yticklabels={0,$ \calibrationintervalinitial $,$ 2\calibrationintervalinitial $},
        xlabel={Time since start of observations, $ t $ (hours)},
        ylabel={Interval (hours)},
        xmin=-10,xmax=416,
        ymin=0,ymax=145,
        legend pos=north west,
        legend columns=4,
        legend image post style={xscale=0.75},
        ]
        % Shade no reference data region
        \draw[fill=gray!25,draw=none] (axis cs:74.83,-200) rectangle (238.56,200);
        \node[anchor=north, inner sep=2pt] () at (axis cs:156.7,50) {No reference data};

        % Initial calibration interval
        \draw[thick, black!60] (axis cs: -100,50) -- (axis cs: 1000,50);

        % Intervals
        \addplot+ [very thick,densely dashed] table[x index=0, y index=1] {data__calibration_schedule_optimisation_20250502112932_intervals.csv};
        \addplot+ [very thick,densely dotted] table[x index=0, y index=2] {data__calibration_schedule_optimisation_20250502112932_intervals.csv};
        \addplot+ [very thick,dashed] table[x index=0, y index=3] {data__calibration_schedule_optimisation_20250502112932_intervals.csv};
        \addplot+ [very thick,dotted] table[x index=0, y index=4] {data__calibration_schedule_optimisation_20250502112932_intervals.csv};

        % Legend
        \legend{Sensor 1, Sensor 2, Sensor 3, Sensor 4}
    \end{axis}
\end{tikzpicture}
    \caption{Adaptive calibration intervals with a 50 hour initial calibration interval.}
    \label{fig:calibration_interval_optimisation}
\end{figure}
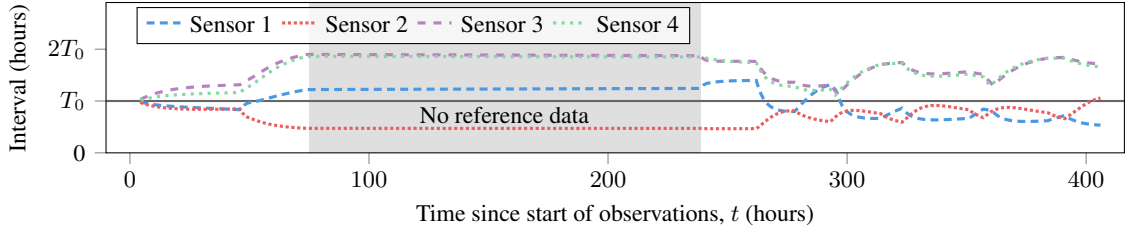
%%%%%%%%%%%%%%%%%%%%%%%%%%%%%%%%%%%%%%%%%%%%%%%%%%%%%%

\cref{fig:calibration_interval_optimisation} shows an example of how the calibration intervals for four sensors can vary over time as they adapt to changes in the prediction uncertainty.
As shown, the calibration interval for sensor 1 is consistently below (shorter than) the initial value, which is due to its higher uncertainty (the underlying sensor data is shown in \cref{fig:sensor_data}).
In contrast, sensors 3 and 4 are consistently above $ \calibrationintervalinitial $ due to their high stability.
%However, observe that the calibration intervals for sensors 3 and 4 fall leading up to 300 hours.
%This is due to the relative stability of sensors 1 and 2 during this time and the higher number of calibrations performed on these sensors, leading to more comparable uncertainty values across all four sensors.
Sudden changes in the graph are driven by calibration events, which, in general, reduce uncertainty for the calibrated sensor.

    \section{Evaluation}
\label{sec:evaluation}

In this section, we first summarise our data collection process and then describe our evaluation of our proposed sensor drift correction and calibration scheduling methods.

\subsection{Data Collection}

%%%%%%%%%%%%%%%%%%%%%%%%%%%%%%%%%%%%%%%%%%%%%%%%%%%
% System overview
\begin{figure}
    \centering
    \footnotesize

\newcommand{\thermostatcolour}{colour2!80!black}
\newcommand{\gascolour}{colour4!70!black}
\newcommand{\sensorscolour}{colour3}
\newcommand{\boxdevice}[4][black!20]{
    \draw[fill=#1] (0,0,0) -- ++(-#2,0,0) -- ++(0,-#3,0) -- ++(#2,0,0) -- cycle;
    \draw[fill=#1] (0,0,0) -- ++(0,0,-#4) -- ++(0,-#3,0) -- ++(0,0,#4) -- cycle;
    \draw[fill=#1] (0,0,0) -- ++(-#2,0,0) -- ++(0,0,-#4) -- ++(#2,0,0) -- cycle;
}

\begin{tikzpicture}[
    Coil/.style={
        thick,
        decoration={#1,segment length=3mm,coil},
        decorate,
    }  % https://tex.stackexchange.com/a/449744
    ]
    \pgfmathsetmacro{\tankx}{2}
    \pgfmathsetmacro{\tanky}{1.5}
    \pgfmathsetmacro{\tankz}{1}

    \pgfmathsetmacro{\airy}{0.3}
    \pgfmathsetmacro{\watery}{1.2}

    \tikzset{
        sensor/.style={minimum width=0.1cm,minimum height=0.45cm,inner sep=0,draw,rounded corners=0.05cm,fill},
    }

    %% Water
    \draw[colour1, fill=colour1!10] (0,-\airy,0) -- ++(-\tankx,0,0) -- ++(0,-\watery,0) -- ++(\tankx,0,0) -- cycle;
    \draw[colour1, fill=colour1!10] (0,-\airy,0) -- ++(0,0,-\tankz) -- ++(0,-\watery,0) -- ++(0,0,\tankz) -- cycle;
    \draw[colour1, fill=colour1!10] (0,-\airy,0) -- ++(-\tankx,0,0) -- ++(0,0,-\tankz) -- ++(\tankx,0,0) -- cycle;

    %% Thermostat and coil
    \node[inner sep=0] (coiltop) at (-0.53\tankx,-0.3,-0.5\tankz) {};
    \draw[Coil={aspect=-0.05,amplitude=-1.5mm},\thermostatcolour,line cap=round] (coiltop.south) -- ++(0,-1.2,0);
    \node[inner sep=0,anchor=south west,\thermostatcolour] (thermostat) at (0.7cm,0.3cm) {
        \begin{tikzpicture}
            \boxdevice[\thermostatcolour!10]{0.5}{0.3}{0.4}
        \end{tikzpicture}
    };
    \draw[rounded corners,\thermostatcolour] (coiltop.south) -- (thermostat -| coiltop) -- (thermostat.west);
    \node[anchor=west,yshift=0.2em] () at (thermostat.east) {Thermostat};

    %% Tank outline
    \draw[thick] (0,0,0) -- ++(-\tankx,0,0) -- ++(0,-\tanky,0) -- ++(\tankx,0,0) -- cycle;
    \draw[thick] (0,0,0) -- ++(0,0,-\tankz) -- ++(0,-\tanky,0) -- ++(0,0,\tankz) -- cycle;
    \draw[thick] (0,0,0) -- ++(-\tankx,0,0) -- ++(0,0,-\tankz) -- ++(\tankx,0,0) -- cycle;
    \draw[thick] (-\tankx,0,-\tankz) -- ++(0,-\airy,0);
    \node[anchor=south east, xshift=-0.4cm, yshift=1.4em] () {Water tank};

    %% Gas mixing system
    \node[inner sep=0,anchor=south,\gascolour] (compressedair) at (-2.7-\tankx,0.2-\tanky,0) {
        \begin{tikzpicture}
            \boxdevice[\gascolour!10]{0.3}{0.8}{0.3}
        \end{tikzpicture}
    };
    \node[inner sep=0,anchor=south,\gascolour] (nitrogen) at (-2.2-\tankx,0.2-\tanky,0) {
        \begin{tikzpicture}
            \boxdevice[\gascolour!10]{0.3}{0.8}{0.3}
        \end{tikzpicture}
    };
    \node[anchor=north] () at ([xshift=-0.1em]compressedair.south) {CA};
    \node[anchor=north] () at (nitrogen.south) {N$_2$};
    \node[inner sep=0,\gascolour] (gasmixer) at (-1.2-\tankx,0.5,0) {
        \begin{tikzpicture}
            \boxdevice[\gascolour!10]{0.5}{0.3}{0.4}
        \end{tikzpicture}
    };
    \node[anchor=north] () at ([xshift=-0.1cm]gasmixer.south) {Gas mixer};
    \draw[rounded corners,double,\gascolour] ([yshift=-0.07cm]compressedair.north) -- ([yshift=-0.01cm]compressedair |- gasmixer) -- ([yshift=-0.01cm]gasmixer.west);
    \draw[rounded corners,double,\gascolour] ([yshift=-0.07cm]nitrogen.north) -- ([yshift=-0.11cm]nitrogen |- gasmixer) -- ([yshift=-0.11cm]gasmixer.west);
    \draw[rounded corners,double,\gascolour] ([xshift=-0.07cm]gasmixer.east) -- ++(1.1,0,0) -- ++(0,-1.85,0) -- ++(0.35,0,0) -- ++(0,0.1,0);

    %% Bubbles
    \node[circle,inner sep=0.080em,draw=\gascolour,line width=0.1pt,fill=white] () at (-0.92*\tankx,-\tanky,-0.5*\tankx) {};
    \node[circle,inner sep=0.075em,draw=\gascolour,line width=0.1pt,fill=white] () at (-0.92*\tankx,-0.9*\tanky,-0.5*\tankx) {};
    \node[circle,inner sep=0.070em,draw=\gascolour,line width=0.1pt,fill=white] () at (-0.92*\tankx,-0.8*\tanky,-0.5*\tankx) {};
    \node[circle,inner sep=0.065em,draw=\gascolour,line width=0.1pt,fill=white] () at (-0.92*\tankx,-0.7*\tanky,-0.5*\tankx) {};

    %% Stirrer
    \begin{pgfonlayer}{background}
        \node[inner sep=0,anchor=north] (stirrer) at (-0.5*\tankx,0.03-\tanky,-0.5*\tankz) {
            \begin{tikzpicture}
                \boxdevice{1.2}{0.1}{1}
            \end{tikzpicture}
        };
    \end{pgfonlayer}
    \draw[latex-latex] ([xshift=-0.5cm,yshift=-0.04cm]stirrer.center) -- ([xshift=0.3cm,yshift=-0.04cm]stirrer.center);
    \node[anchor=west,inner sep=0] (stirrerlabel) at ([xshift=-0.15cm,yshift=-0.16cm]stirrer.east) {Stirrer};

    %% Electrochemical sensors
    \node[sensor,\sensorscolour] (sensor1) at (-0.9*\tankx,-0.7*\tanky,-0.5*\tankz) {};
    \node[sensor,\sensorscolour] (sensor2) at (-0.75*\tankx,-0.7*\tanky,-0.5*\tankz) {};
    \node[sensor,\sensorscolour] (sensor3) at (-0.6*\tankx,-0.7*\tanky,-0.5*\tankz) {};
    \node[sensor,\sensorscolour] (sensor4) at (-0.45*\tankx,-0.7*\tanky,-0.5*\tankz) {};
    \node[anchor=north east] (sensorslabel) at ([xshift=-1.2cm,yshift=-0.63cm]sensor1.south) {EC sensors};
    \draw[->,thick,densely dashed,\sensorscolour] ([xshift=0cm]sensorslabel.east) to [bend right=19] ([xshift=-0.1cm,yshift=-0.1cm]sensor3.south west);

    %% Data logger 1
    \node[inner sep=0,\sensorscolour] (datalogger1) at (-0.7-\tankx,-0.8*\tanky,-0.5*\tankz) {
        \begin{tikzpicture}
            \boxdevice[\sensorscolour!10]{0.5}{0.3}{0.3}
        \end{tikzpicture}
    };
    \node[anchor=east,align=right,yshift=-0.2em,xshift=-0.2em] () at (datalogger1.west) {Data\\logger};

    %% Wires from sensors
    \draw[rounded corners,\sensorscolour] (sensor1.north) -- ([yshift=0.9cm]sensor1.north) -- ([yshift=0.9cm]sensor1.north -| datalogger1) -- ([yshift=-0.07cm]datalogger1.north);
    \draw[rounded corners,\sensorscolour] (sensor2.north) -- ([yshift=0.9cm]sensor2.north) -- ([yshift=0.9cm]sensor2.north -| datalogger1) -- ([yshift=-0.07cm]datalogger1.north);
    \draw[rounded corners,\sensorscolour] (sensor3.north) -- ([yshift=0.9cm]sensor3.north) -- ([yshift=0.9cm]sensor3.north -| datalogger1) -- ([yshift=-0.07cm]datalogger1.north);
    \draw[rounded corners,\sensorscolour] (sensor4.north) -- ([yshift=0.9cm]sensor4.north) -- ([yshift=0.9cm]sensor4.north -| datalogger1) -- ([yshift=-0.07cm]datalogger1.north);

    %% Data logger 2
    \node[inner sep=0,\sensorscolour] (datalogger2) at (0.8,-0.8*\tanky,-0.5*\tankz) {
        \begin{tikzpicture}
            \boxdevice[\sensorscolour!10]{0.5}{0.3}{0.3}
        \end{tikzpicture}
    };
    \node[anchor=west,align=left,yshift=-0.1em,xshift=0.1em] () at (datalogger2.east) {Data\\logger};

    %% Optical sensor
    \node[inner sep=0,anchor=west,\sensorscolour] (optical) at (0,-0.5*\tanky,-0.5*\tankz) {
        \begin{tikzpicture}
            \boxdevice[\sensorscolour!10]{0.2}{0.3}{0.3}
        \end{tikzpicture}
    };
    \node[anchor=west,align=left,yshift=0,xshift=0] () at (optical.north east) {Optical sensor};
    \draw[rounded corners,\sensorscolour] ([xshift=-0.05cm]optical.east) -- (optical -| datalogger2) -- ([yshift=-0.07cm]datalogger2.north);

    %% PC
    \node[inner sep=0,anchor=west] (pcbase) at ([xshift=1.2cm]datalogger2.south east) {
        \begin{tikzpicture}
            \boxdevice{0.8}{0.05}{0.6}
        \end{tikzpicture}
    };
    \node[inner sep=0,anchor=south] (pcscreen) at ([xshift=0.12cm,yshift=-0.01cm]pcbase.north) {
        \begin{tikzpicture}
            \boxdevice{0.8}{0.55}{0.02}
        \end{tikzpicture}
    };
    \draw[] ([xshift=0.07cm,yshift=0.07cm]pcscreen.south west)
    -- ([xshift=-0.07cm,yshift=0.07cm]pcscreen.south east)
    -- ([xshift=-0.07cm,yshift=-0.07cm]pcscreen.north east)
    -- ([xshift=0.07cm,yshift=-0.07cm]pcscreen.north west) -- cycle;
    \node[anchor=north west] () at ([xshift=-0.09cm]pcbase.south west) {PC};

    %% Wires to PC
    \draw[rounded corners,\sensorscolour] (datalogger1.south) -- ([yshift=-0.65cm]datalogger1.south |- pcbase.south) -- ([yshift=-0.65cm]pcbase.south) -- (pcbase.south);
    \draw[rounded corners,\sensorscolour] (datalogger2.south) -- ([yshift=-0.65cm]datalogger2.south |- pcbase.south) -- ([yshift=-0.65cm]pcbase.south) -- (pcbase.south);
    \draw[rounded corners,\gascolour] ([yshift=-0.07cm]gasmixer.north) -- ([yshift=0.3cm]gasmixer.north) -- ([yshift=0.3cm]gasmixer.north -| pcscreen) -- (pcscreen.north);

\end{tikzpicture}
    \caption{Experimental setup with sensor components shown in \textcolor{\sensorscolour}{purple} and gas mixing components in \textcolor{\gascolour}{green}. EC~=~electrochemical. CA = compressed air.}
    \label{fig:data_collection}
\end{figure}
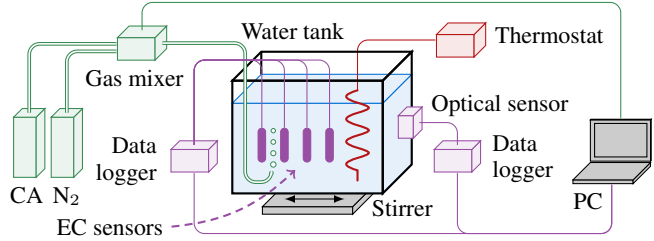
%%%%%%%%%%%%%%%%%%%%%%%%%%%%%%%%%%%%%%%%%%%%%%%%%%%

%%%%%%%%%%%%%%%%%%%%%%%%%%%%%%%%%%%%%%%%%%%%%%%%%%%
% Sensor data
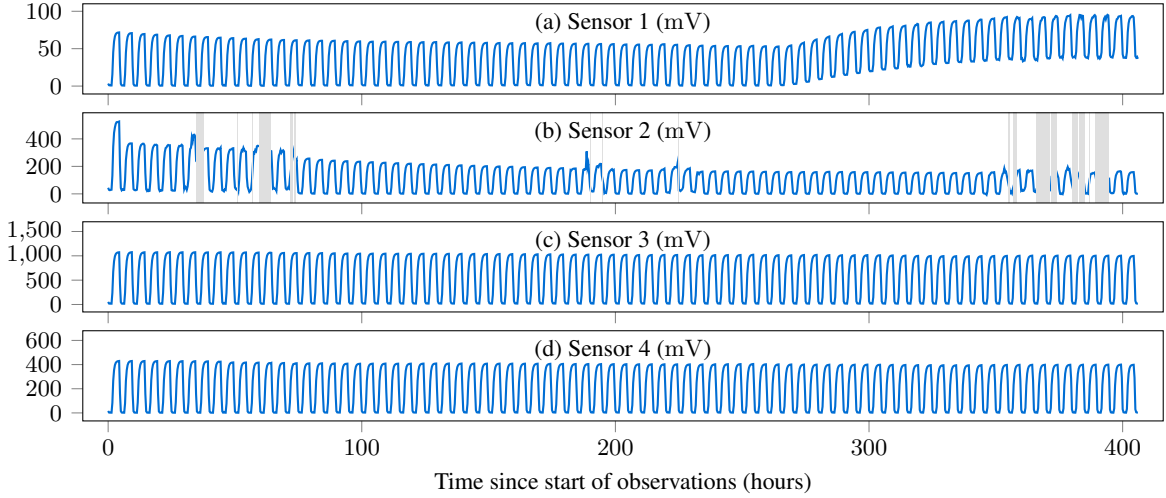
\begin{figure}[!t]
    \centering
    \begin{tikzpicture}
    \begin{groupplot}[
        group style={rows=4,xlabels at=edge bottom,vertical sep=0.25cm,xticklabels at=edge bottom},
        generic line plot,
        height=1.2cm,
        width=0.87\columnwidth,
        xtick={0,100,200,300,400},
        xlabel={Time since start of observations (hours)},
        xmin=-10,xmax=416,
        ]
        \nextgroupplot[ymax=110]
            \addplot [thick,colour1] table[x index=0, y index=1] {data__sensor_data_sparser.csv};
            \node[anchor=north,font=\footnotesize] () at (axis cs:203,110) {(a) Sensor 1 (\unit{mV})};
        \nextgroupplot[ymax=590]
            \addplot [thick,colour1] table[x index=0, y index=2] {data__sensor_data_sparser.csv};
            \sensorTwoMissingData
            \node[anchor=north,font=\footnotesize] () at (axis cs:203,590) {(b) Sensor 2 (\unit{mV})};
        \nextgroupplot[ymax=1700]
            \addplot [thick,colour1] table[x index=0, y index=3] {data__sensor_data_sparser.csv};
            \node[anchor=north,font=\footnotesize] () at (axis cs:203,1700) {(c) Sensor 3 (\unit{mV})};
        \nextgroupplot[ymax=680]
            \addplot [thick,colour1] table[x index=0, y index=4] {data__sensor_data_sparser.csv};
            \node[anchor=north,font=\footnotesize] () at (axis cs:203,680) {(d) Sensor 4 (\unit{mV})};
    \end{groupplot}
\end{tikzpicture}
    \caption{
        Data collected from each electrochemical sensor.
        Invalid readings from sensor 2 are indicated by grey regions.
    }
    \label{fig:sensor_data}
\end{figure}
%%%%%%%%%%%%%%%%%%%%%%%%%%%%%%%%%%%%%%%%%%%%%%%%%%%

Data for evaluating our approach was collected from four electrochemical DO sensors placed in a tank containing water at a constant temperature of 25.36$\pm$0.16\textdegree{}C.
In total, 12,186 samples were collected per sensor at two minute intervals over 406 hours (approximately 17 days).
During this period, the tank was alternately bubbled with compressed air for three hours, saturating the water, followed by two hours of nitrogen gas (N$_2$), creating a purely anoxic environment.
The overall data collection setup is illustrated in \cref{fig:data_collection}.

\cref{fig:sensor_data} displays the data collected from each sensor, clearly illustrating the oscillating cycles between saturation and de-saturation.
There is a general trend across all sensors such that the height of cyclical peaks decreases.
This indicates that sensor drift is present in the data and, specifically, corresponds to a reduction in sensitivity ($ \coefficient_1 $).
However, importantly, the degree and nature of sensor drift varies significantly between the sensors.
Sensor 2 also experienced periods of invalid readings, which have been removed in subsequent analysis (indicated by grey regions in \cref{fig:sensor_data}(b)).

Dissolved oxygen, temperature, humidity and air pressure were also monitored using a reference optical sensor (Firesting), which is considered the ground truth.
Note that the Firesting sensor was unavailable between 75 and 238 hours after the start of observations (approximately 40\% of the duration).

\begin{figure}[!t]
    \centering
    \begin{tikzpicture}
    \begin{axis}[
        generic line plot,
        width=0.8\columnwidth,
        height=1.7cm,
        xtick={269.6, 269.7, 269.8, 269.9, 270.0, 270.1},
        ytick={0, 0.2, 0.4, 0.6, 0.8, 1},
        xmin=269.5,xmax=270.18,
        xlabel={Time since start of observations (hours)},
        ylabel={Normalised},
        ]
        %% Draw plots
        \addplot+ [no marks,black] table[x index=0, y index=1] {data__timing_correction_sensor_3.csv};
        \addplot+ [no marks,densely dashed,ultra thick] table[x index=0, y index=1] {data__timing_correction_ref_original.csv};
        \addplot+ [no marks,dashed,ultra thick,colour4] table[x index=0, y index=1] {data__timing_correction_ref_adjusted.csv};

        %% Legend
        \legend{Sensor 3, Firesting (original), Firesting (sync)}

        %% Annotations
        \node[font=\small,anchor=north west] () at (axis cs:269.5,0.95) {(a)};
%        \draw[red,-latex,very thick] (axis cs:269.56,0.7) -- (axis cs:269.635,0.7);
%        \draw[red,latex-,very thick] (axis cs:269.677,0.7) -- (axis cs:269.75,0.7);
%        \node[red,anchor=east,text depth=0.3em,text height=0.8em] () at (axis cs:269.56,0.7) {Timing offset};
    \end{axis}
\end{tikzpicture}
    \begin{tikzpicture}
    \begin{axis}[
        generic scatter plot,
        width=0.8\columnwidth,
        height=1.7cm,
        ymin=-120,ymax=120,
        xtick={0,100,200,300,400},
        xmin=-10,xmax=408,
        xlabel={Time since start of observations (hours)},
        ylabel={$ \tau_{\text{\tiny offset}} $ (sec)},
        ]
        %% Draw plot
        \addplot+ [line width=1,mark=x,mark options={scale=1.0}] table[x index=0, y index=1] {data__timing_correction_best_offsets.csv};

        %% Shade no data region
        \draw[fill=gray!25,draw=none] (axis cs:74.83,-200) rectangle (238.56,200);
        \node[anchor=center] () at (axis cs:156.7,-20) {No Firesting data};

        %% Annotations
        \node[font=\small,anchor=north west] () at (axis cs:-10,120) {(b)};
    \end{axis}
\end{tikzpicture}
    \caption{
        Signal timing offset correction (a) at one saturation-anoxic transition and (b) the optimal offset over time.
    }
    \label{fig:timing_correction}
\end{figure}
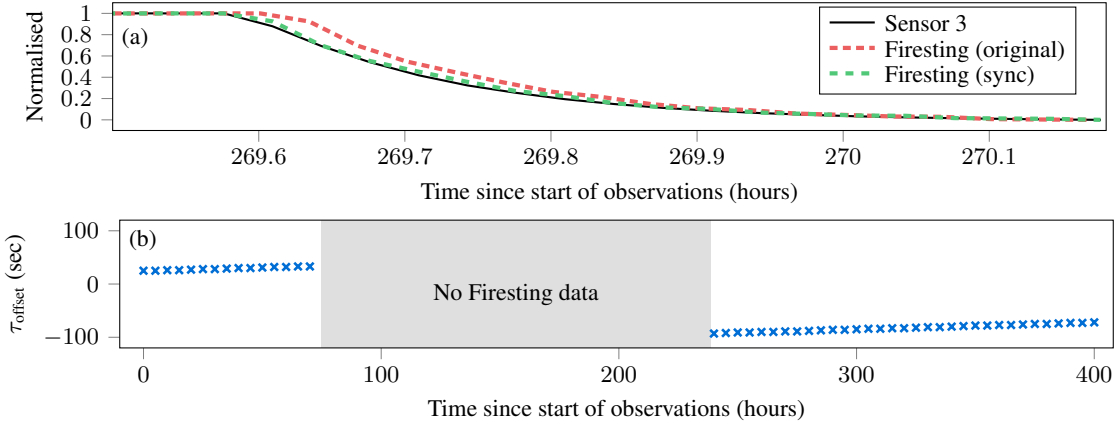

As illustrated in \cref{fig:timing_correction}(a), which shows a single saturation-anoxic transition, the clocks for the Firesting and electrochemical sensors data loggers were not perfectly synchronised.
To rectify this, we identified the offsets (between $ \pm $200 seconds) that resulted in maximum correlation between the two signals at five hour intervals.
The correlation-maximising offsets are shown in \cref{fig:timing_correction}(b) and indicate that the Firesting data logger's clock loses approximately \qty{0.129}{s} per hour relative to the electrochemical sensors.
Synchronisation was thus achieved by adding the following offset to the Firesting timestamps:
\begin{equation}
    \tau_{\mathrm{offset}} =
    \begin{cases}
        0.129 t + 24.5  & t \le 100 \\
        0.129 t - 123.9 & t > 100
    \end{cases}
\end{equation}
in seconds, where $ t $ is the number of hours since the start of observations.
After synchronisation, there is still a small delay in the Firesting data due to the slower response time of optical DO sensors.

\subsection{Sensor Drift Correction}
\label{sec:evaluation:driftcorrection}

%We tested our proposed GPR-based sensor drift correction approach across a wide range of calibration frequencies.
Calibrations were obtained by selecting 3--6 temporally aligned samples from the electrochemical and optical (reference) sensors at low (near-0\%) and high (near-100\%) dissolved air levels.
The two sets of samples are collected within a single saturation-anoxic cycle to ensure close temporal proximity and minimal intra-calibration drift.
Linear regression is then performed on these samples to estimate the response function coefficients and their standard errors.
We tested calibration intervals, i.e., the mean time between calibrations, from 10 to 100 hours in 10 hour increments.
For each interval, 50 repetitions were performed for each sensor with the first calibration performed at a random time between zero and the lesser of 50 hours or the calibration interval length.
Three GPR kernels were tested, namely RBF, RQ and Matérn (see \cref{sec:background:gpr}), with a range of kernel lengths from 50 to 400 hours.

%%%%%%%%%%%%%%%%%%%%%%%%%%%%%%%%%%%%%%%%%%%%%%%%%%%
% Interpolation methods & results for each kernel
\begin{figure}[!t]
    \centering
    \begin{tikzpicture}
    \begin{axis}[
        generic scatter plot,
        width=0.8\columnwidth,
        height=2cm,
        ymin=-8,ymax=52,
        xtick={0,100,200,300,400},
        xmin=-10,xmax=416,
        xlabel={Time since start of observations (hours)},
        ylabel={Intercept, $ \coefficient_0 $},
        ]
        % Calibrations
        \addplot [black,only marks,mark size=1.5pt,mark=*] table[x index=0, y index=1] {data__sensor_2_10_hrs_matern_length_200_calibrations.csv};

        % Shade no reference data region
        \draw[fill=gray!25,draw=none] (axis cs:74.83,-200) rectangle (238.56,200);
        \node[anchor=north] () at (axis cs:156.7,50) {No reference data};

        % Stepwise
        \addplot [const plot, no marks, colour1, very thick] table[x index=0, y index=1] {data__sensor_2_10_hrs_matern_length_200_calibrations.csv};
        \draw[colour1, very thick] (axis cs: 0,25.819230355953717) -- (axis cs: 4.583472222222222,25.819230355953717);
        \draw[colour1, very thick] (axis cs: 420,4.318850189530405) -- (axis cs: 397.1141666666666,4.318850189530405);

        % Linear
        \addplot [sharp plot, no marks, colour2, densely dashed, very thick] table[x index=0, y index=1] {data__sensor_2_10_hrs_matern_length_200_calibrations.csv};
        \draw[colour2, densely dashed, very thick] (axis cs: 0,25.819230355953717) -- (axis cs: 4.583,25.8192);
        \draw[colour2, densely dashed, very thick] (axis cs: 420,4.318850189530405) -- (axis cs: 397.1141,4.3188);

        \legend{Calibrations, Stepwise, Linear}
    \end{axis}
\end{tikzpicture}
    \caption{
        Different interpolation methods applied to data from \cref{fig:example}(b).
    }
    \label{fig:interpolation_methods}
\end{figure}
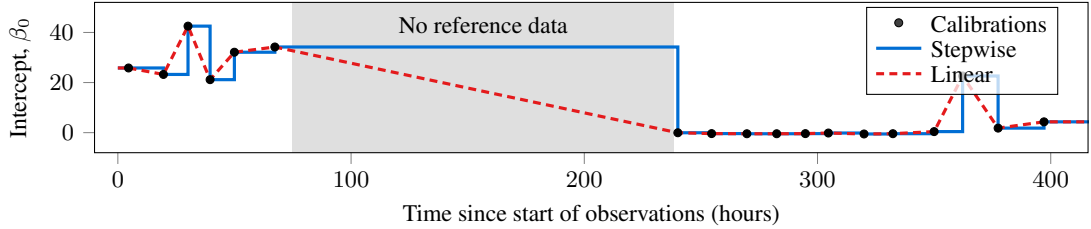

\begin{table}
    \centering
    \caption{
        Median MSE for each kernel type and length.  % across all sensors and interval lengths.
    }
    \footnotesize

\begin{tabular}{l|rrrr|rrrr}
    \toprule
           &   \multicolumn{4}{c|}{\textbf{Offline}}   & \multicolumn{4}{|c}{\textbf{Online}}       \\
    Kernel &           50 &  100 &          200 &  400 &            50 &  100 &          200 &  400 \\ \midrule
    RBF    & \textbf{7.8} & 13.1 &         10.5 & 14.3 & \textbf{70.7} & 73.4 &         82.7 & 97.7 \\
    RQ     &          2.6 &  2.6 & \textbf{2.5} &  2.6 &           8.6 &  6.7 & \textbf{6.0} &  6.4 \\
    Matérn &          2.4 &  2.4 & \textbf{2.3} &  2.4 &           6.7 &  5.7 & \textbf{3.9} &  6.4 \\ \bottomrule
\end{tabular}
    \label{tab:kernel_length}
\end{table}
%%%%%%%%%%%%%%%%%%%%%%%%%%%%%%%%%%%%%%%%%%%%%%%%%%%

\subsubsection{Optimal Kernel Lengths}

First, we examine the effect of kernel length on performance.
\cref{tab:kernel_length} shows the median MSE for each kernel type and length evaluated across all sensors and interval lengths for both offline and online scenarios.
As can be seen, a kernel length of 50~hours is most suitable for the RBF kernel, while 200~hours delivers optimal performance when using either RQ and Matérn kernels.
This is clearly a very high-level evaluation that hides a significant amount of variation.
However, we find that the conclusion regarding optimal kernel lengths holds true in the vast majority of our experiments or otherwise results in very little difference.
We therefore continue with these parameter values for the remainder of our evaluation.

\subsubsection{Offline Drift Correction}

%%%%%%%%%%%%%%%%%%%%%%%%%%%%%%%%%%%%%%%%%%%%%%%%%%%
% Offline results figure
\begin{figure*}
    \centering
    \begin{tikzpicture}
    \begin{groupplot}[
        group style={group size=2 by 2,xlabels at=edge bottom,vertical sep=0.5cm,horizontal sep=1.5cm,xticklabels at=edge bottom},
        generic line plot,
        width=0.42\columnwidth,
        height=2.5cm,
        xmin=5,xmax=105,
        ymin=0,ymax=1.28,
        xtick={0, 20, 40, 60, 80, 100},
        ytick={0,0.2,0.4,0.6,0.8,1,1.2},
        restrict y to domain=0:100,  % avoid "Dimension too large issue with RBF values
        xlabel={Calibration interval (hours)},
        ylabel={Relative MSE},
        legend columns=2,
        legend pos=south east,
        legend style={tikz/column 2/.style={column sep=8pt}}
        ]
        %%%%%%%%%%%%%%%%%%%%%%%%%%%%%%%%%%%%%%%%%%%%%%%%%%%%%%
        % Sensor 1
        \nextgroupplot[legend pos=north west]
            % Fill good region
            \draw[colour4!10, fill] (axis cs:5,0) -- (axis cs:105,0) -- (axis cs:105,1) -- (axis cs:5,1) -- cycle;

            % Draw plots
            \addplot [black, thick, mark=diamond*] table[x index=0, y index=1] {data__drift_correction_offline_results_norm.csv};
            \pgfplotsset{cycle list shift=-1}
            \addplot+ [color=colour2, thick] table[x index=0, y index=2] {data__drift_correction_offline_results_norm.csv};
            \addplot+ [color=colour1, thick] table[x index=0, y index=3] {data__drift_correction_offline_results_norm.csv};
            \addplot+ [color=colour3, thick] table[x index=0, y index=4] {data__drift_correction_offline_results_norm.csv};

            % Annotate reference line
            \draw[densely dashed, thick, black!50] (axis cs:-100,1) -- (axis cs:1000,1);

            %% Label
            \node[font=\footnotesize,anchor=north east] () at (axis cs:105,1.25) {(a) Sensor 1};

            %% Legend
            \legend{Linear, RBF-50, RQ-200, Matérn-200}

        %%%%%%%%%%%%%%%%%%%%%%%%%%%%%%%%%%%%%%%%%%%%%%%%%%%%%%
        % Sensor 2
        \nextgroupplot
            % Fill good region
            \draw[colour4!10, fill] (axis cs:5,0) -- (axis cs:105,0) -- (axis cs:105,1) -- (axis cs:5,1) -- cycle;

            %% Draw plots
            \addplot [black, thick, mark=diamond*] table[x index=0, y index=5] {data__drift_correction_offline_results_norm.csv};
            \pgfplotsset{cycle list shift=-1}
            \addplot+ [color=colour2, thick] table[x index=0, y index=6] {data__drift_correction_offline_results_norm.csv};
            \addplot+ [color=colour1, thick] table[x index=0, y index=7] {data__drift_correction_offline_results_norm.csv};
            \addplot+ [color=colour3, thick] table[x index=0, y index=8] {data__drift_correction_offline_results_norm.csv};

            % Annotate reference line
            \draw[densely dashed, thick, black!50] (axis cs:-100,1) -- (axis cs:1000,1);

            %% Label
            \node[font=\footnotesize,anchor=north east] () at (axis cs:105,1.25) {(b) Sensor 2};

            %% Legend
            \legend{Linear, RBF-50, RQ-200, Matérn-200}

        %%%%%%%%%%%%%%%%%%%%%%%%%%%%%%%%%%%%%%%%%%%%%%%%%%%%%%
        % Sensor 3
        \nextgroupplot
            % Fill good region
            \draw[colour4!10, fill] (axis cs:5,0) -- (axis cs:105,0) -- (axis cs:105,1) -- (axis cs:5,1) -- cycle;

            %% Draw plots
            \addplot [black, thick, mark=diamond*] table[x index=0, y index=9] {data__drift_correction_offline_results_norm.csv};
            \pgfplotsset{cycle list shift=-1}
            \addplot+ [color=colour2, thick] table[x index=0, y index=10] {data__drift_correction_offline_results_norm.csv};
            \addplot+ [color=colour1, thick] table[x index=0, y index=11] {data__drift_correction_offline_results_norm.csv};
            \addplot+ [color=colour3, thick] table[x index=0, y index=12] {data__drift_correction_offline_results_norm.csv};

            % Annotate reference line
            \draw[densely dashed, thick, black!50] (axis cs:-100,1) -- (axis cs:1000,1);

            %% Label
            \node[font=\footnotesize,anchor=north east] () at (axis cs:105,1.25) {(c) Sensor 3};

            %% Legend
            \legend{Linear, RBF-50, RQ-200, Matérn-200}

        %%%%%%%%%%%%%%%%%%%%%%%%%%%%%%%%%%%%%%%%%%%%%%%%%%%%%%
        % Sensor 4
        \nextgroupplot
            % Fill good region
            \draw[colour4!10, fill] (axis cs:5,0) -- (axis cs:105,0) -- (axis cs:105,1) -- (axis cs:5,1) -- cycle;

            %% Draw plots
            \addplot [black, thick, mark=diamond*] table[x index=0, y index=13] {data__drift_correction_offline_results_norm.csv};
            \pgfplotsset{cycle list shift=-1}
            \addplot+ [color=colour2, thick] table[x index=0, y index=14] {data__drift_correction_offline_results_norm.csv};
            \addplot+ [color=colour1, thick] table[x index=0, y index=15] {data__drift_correction_offline_results_norm.csv};
            \addplot+ [color=colour3, thick] table[x index=0, y index=16] {data__drift_correction_offline_results_norm.csv};

            % Annotate reference line
            \draw[densely dashed, thick, black!50] (axis cs:-100,1) -- (axis cs:1000,1);

            %% Label
            \node[font=\footnotesize,anchor=north east] () at (axis cs:105,1.25) {(d) Sensor 4};

            %% Legend
            \legend{Linear, RBF-50, RQ-200, Matérn-200}

    \end{groupplot}
\end{tikzpicture}
    \caption{Offline drift correction: median relative MSE for each sensor at different calibration intervals.}
    \label{fig:drift_correction_offline}
\end{figure*}
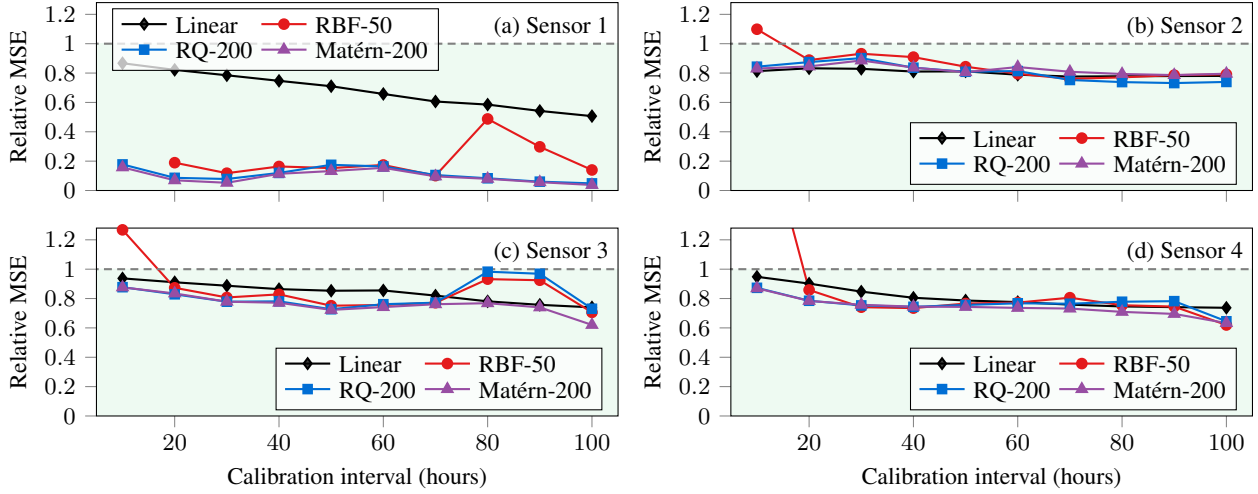
%%%%%%%%%%%%%%%%%%%%%%%%%%%%%%%%%%%%%%%%%%%%%%%%%%%

Offline drift correction is tested by training the GPR models on all calibrations.
Predictions at time $ t $ are thus made using models that are trained using both past and future calibrations.
For offline drift correction we compare to stepwise and linear interpolation, which are illustrated in \cref{fig:interpolation_methods} and involve either carrying the most recent calibration forward until the next calibration or using a linear interpolation between calibrations, respectively.
Performance is evaluated using relative MSE compared to stepwise interpolation, which is calculated by dividing each method's MSE by the MSE for stepwise interpolation.
As such, a value of 1 corresponds to no change relative to stepwise interpolation and lower values indicate improved accuracy.

As shown in \cref{fig:drift_correction_offline} and \cref{tab:drift_correction_offline}, our approach is consistently able to reduce the MSE for each sensor.
In particular, the Matérn kernel proves to be the most effective for drift correction with generally lower relative MSE values.
Sensor 1 exhibits the greatest MSE reductions due to drift correction using our approach, with the Matérn kernel able to reduce MSE by over 90\% on average.
For the remaining sensors, the improvement is typically 10--30\% and approximately 23\% on average for the Matérn kernel.
Notably, this also applies to sensors 3 and 4, which experience minimal drift, suggesting that our approach is still effective even for small sensor drift.
Our proposed GPR-based drift correction method also comfortably outperforms linear interpolation for three out of four sensors, with sensor 2 being the lone exception.
This is likely due to the high degree of noise exhibited by this sensor, which outweighs sensor drift in terms of contribution to MSE.

%%%%%%%%%%%%%%%%%%%%%%%%%%%%%%%%%%%%%%%%%%%%%%%%%%%
% Offline results table
\begin{table}
    \centering
    \caption{
        Relative MSE for offline drift correction for each sensor and calibration interval.
        The best performance for each sensor and interval is highlighted in \textbf{bold}.
    }
    \footnotesize

\setlength{\tabcolsep}{0.3em}

\begin{tabular}{r|rrrr|rrrr|rrrr|rrrr}
    \toprule
             & \multicolumn{4}{c|}{\textbf{Sensor 1}} &        \multicolumn{4}{c|}{\textbf{Sensor 2}}         &    \multicolumn{4}{c|}{\textbf{Sensor 3}}     &     \multicolumn{4}{c}{\textbf{Sensor 4}}     \\
    Interval & Linear &    RBF &   RQ &        Matérn &        Linear &  RBF &            RQ &         Matérn & Linear &  RBF &            RQ &        Matérn & Linear &           RBF &   RQ &        Matérn \\ \midrule
          10 &   0.87 & 512.73 & 0.18 & \textbf{0.16} & \textbf{0.81} & 1.10 &          0.84 &           0.83 &   0.94 & 1.27 &          0.88 & \textbf{0.88} &   0.95 &          2.19 & 0.87 & \textbf{0.87} \\
          20 &   0.82 &   0.19 & 0.09 & \textbf{0.07} & \textbf{0.83} & 0.89 &          0.87 &           0.85 &   0.91 & 0.87 & \textbf{0.83} &          0.83 &   0.90 &          0.86 & 0.78 & \textbf{0.78} \\
          30 &   0.78 &   0.12 & 0.08 & \textbf{0.05} & \textbf{0.83} & 0.93 &          0.90 &           0.89 &   0.89 & 0.81 &          0.78 & \textbf{0.78} &   0.85 & \textbf{0.74} & 0.75 &          0.76 \\
          40 &   0.75 &   0.16 & 0.12 & \textbf{0.11} & \textbf{0.81} & 0.91 &          0.84 &           0.84 &   0.86 & 0.83 &          0.78 & \textbf{0.77} &   0.81 & \textbf{0.73} & 0.74 &          0.74 \\
          50 &   0.71 &   0.15 & 0.17 & \textbf{0.13} &          0.81 & 0.84 &          0.81 & \textbf{ 0.81} &   0.85 & 0.75 &          0.73 & \textbf{0.72} &   0.79 &          0.77 & 0.76 & \textbf{0.74} \\
          60 &   0.66 &   0.17 & 0.16 & \textbf{0.15} & \textbf{0.79} & 0.80 &          0.81 &           0.84 &   0.86 & 0.76 &          0.76 & \textbf{0.74} &   0.78 &          0.77 & 0.77 & \textbf{0.74} \\
          70 &   0.61 &   0.10 & 0.11 & \textbf{0.10} &          0.78 & 0.76 & \textbf{0.75} &           0.81 &   0.82 & 0.77 &          0.77 & \textbf{0.76} &   0.76 &          0.81 & 0.76 & \textbf{0.73} \\
          80 &   0.59 &   0.49 & 0.08 & \textbf{0.08} &          0.78 & 0.77 & \textbf{0.74} &           0.79 &   0.78 & 0.93 &          0.98 & \textbf{0.77} &   0.75 &          0.75 & 0.78 & \textbf{0.71} \\
          90 &   0.54 &   0.30 & 0.06 & \textbf{0.06} &          0.78 & 0.79 & \textbf{0.73} &           0.79 &   0.76 & 0.92 &          0.97 & \textbf{0.74} &   0.74 &          0.75 & 0.78 & \textbf{0.70} \\
         100 &   0.51 &   0.14 & 0.05 & \textbf{0.04} &          0.78 & 0.79 & \textbf{0.74} &           0.80 &   0.74 & 0.71 &          0.73 & \textbf{0.62} &   0.74 & \textbf{0.62} & 0.65 &          0.63 \\ \midrule
         All &   0.69 &   0.18 & 0.10 & \textbf{0.09} &          0.81 & 0.84 & \textbf{0.80} &           0.83 &   0.85 & 0.81 &          0.80 & \textbf{0.76} &   0.81 &          0.77 & 0.77 & \textbf{0.73} \\ \bottomrule
\end{tabular}

    \label{tab:drift_correction_offline}
\end{table}
%%%%%%%%%%%%%%%%%%%%%%%%%%%%%%%%%%%%%%%%%%%%%%%%%%%

\subsubsection{Online Drift Correction}

%%%%%%%%%%%%%%%%%%%%%%%%%%%%%%%%%%%%%%%%%%%%%%%%%%%
% Online results
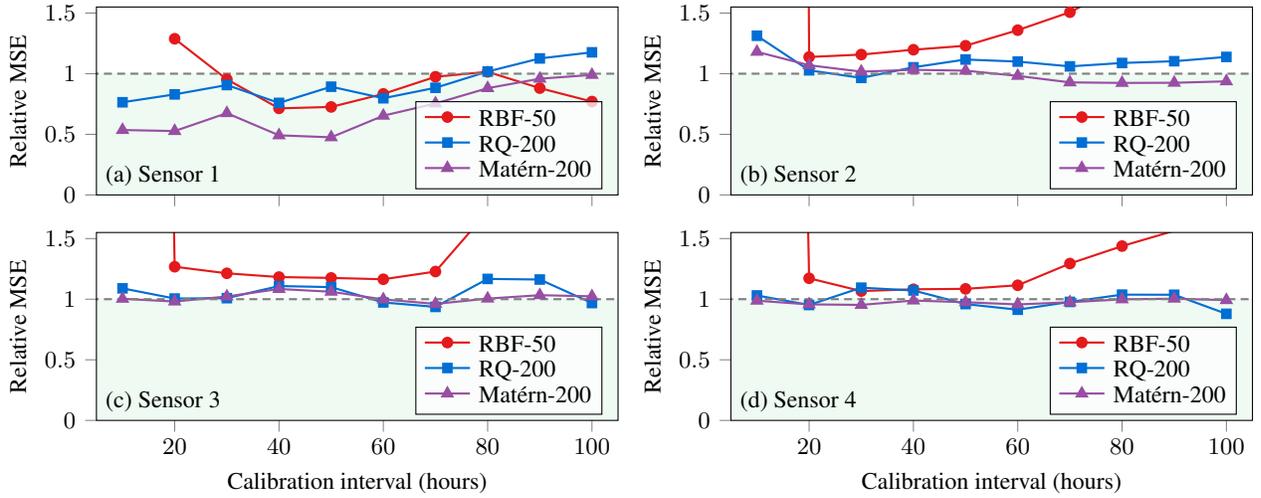
\begin{figure*}
    \centering
    \begin{tikzpicture}
    \begin{groupplot}[
        group style={group size=2 by 2,xlabels at=edge bottom,vertical sep=0.5cm,horizontal sep=1.5cm,xticklabels at=edge bottom},
        generic line plot,
        width=0.42\columnwidth,
        height=2.5cm,
        xmin=5,xmax=105,
        ymin=0,ymax=1.55,
        xtick={0, 20, 40, 60, 80, 100},
        restrict y to domain=0:100,  % avoid "Dimension too large issue with RBF values
        xlabel={Calibration interval (hours)},
        ylabel={Relative MSE},
        legend columns=1,
        legend pos=south east,
        legend style={tikz/column 2/.style={column sep=8pt}}
        ]
        %%%%%%%%%%%%%%%%%%%%%%%%%%%%%%%%%%%%%%%%%%%%%%%%%%%%%%
        % Sensor 1
        \nextgroupplot
            % Fill good region
            \draw[colour4!10, fill] (axis cs:5,0) -- (axis cs:105,0) -- (axis cs:105,1) -- (axis cs:5,1) -- cycle;

            % Draw plots
            \addplot+ [color=colour2, thick] table[x index=0, y index=1] {data__drift_correction_online_results_norm.csv};
            \addplot+ [color=colour1, thick] table[x index=0, y index=2] {data__drift_correction_online_results_norm.csv};
            \addplot+ [color=colour3, thick] table[x index=0, y index=3] {data__drift_correction_online_results_norm.csv};

            % Annotate reference line
            \draw[densely dashed, thick, black!50] (axis cs:-100,1) -- (axis cs:1000,1);

            %% Label
            \node[font=\footnotesize,anchor=south west] () at (axis cs:5,0) {(a) Sensor 1};

            %% Legend
            \legend{RBF-50, RQ-200, Matérn-200}

        %%%%%%%%%%%%%%%%%%%%%%%%%%%%%%%%%%%%%%%%%%%%%%%%%%%%%%
        % Sensor 2
        \nextgroupplot
            % Fill good region
            \draw[colour4!10, fill] (axis cs:5,0) -- (axis cs:105,0) -- (axis cs:105,1) -- (axis cs:5,1) -- cycle;

            %% Draw plots
            \addplot+ [color=colour2, thick] table[x index=0, y index=4] {data__drift_correction_online_results_norm.csv};
            \addplot+ [color=colour1, thick] table[x index=0, y index=5] {data__drift_correction_online_results_norm.csv};
            \addplot+ [color=colour3, thick] table[x index=0, y index=6] {data__drift_correction_online_results_norm.csv};

            % Annotate reference line
            \draw[densely dashed, thick, black!50] (axis cs:-100,1) -- (axis cs:1000,1);

            %% Label
            \node[font=\footnotesize,anchor=south west] () at (axis cs:5,0) {(b) Sensor 2};

            %% Legend
            \legend{RBF-50, RQ-200, Matérn-200}

        %%%%%%%%%%%%%%%%%%%%%%%%%%%%%%%%%%%%%%%%%%%%%%%%%%%%%%
        % Sensor 3
        \nextgroupplot
            % Fill good region
            \draw[colour4!10, fill] (axis cs:5,0) -- (axis cs:105,0) -- (axis cs:105,1) -- (axis cs:5,1) -- cycle;

            %% Draw plots
            \addplot+ [color=colour2, thick] table[x index=0, y index=7] {data__drift_correction_online_results_norm.csv};
            \addplot+ [color=colour1, thick] table[x index=0, y index=8] {data__drift_correction_online_results_norm.csv};
            \addplot+ [color=colour3, thick] table[x index=0, y index=9] {data__drift_correction_online_results_norm.csv};

            % Annotate reference line
            \draw[densely dashed, thick, black!50] (axis cs:-100,1) -- (axis cs:1000,1);

            %% Label
            \node[font=\footnotesize,anchor=south west] () at (axis cs:5,0) {(c) Sensor 3};

            %% Legend
            \legend{RBF-50, RQ-200, Matérn-200}

        %%%%%%%%%%%%%%%%%%%%%%%%%%%%%%%%%%%%%%%%%%%%%%%%%%%%%%
        % Sensor 4
        \nextgroupplot
            % Fill good region
            \draw[colour4!10, fill] (axis cs:5,0) -- (axis cs:105,0) -- (axis cs:105,1) -- (axis cs:5,1) -- cycle;

            %% Draw plots
            \addplot+ [color=colour2, thick] table[x index=0, y index=10] {data__drift_correction_online_results_norm.csv};
            \addplot+ [color=colour1, thick] table[x index=0, y index=11] {data__drift_correction_online_results_norm.csv};
            \addplot+ [color=colour3, thick] table[x index=0, y index=12] {data__drift_correction_online_results_norm.csv};

            % Annotate reference line
            \draw[densely dashed, thick, black!50] (axis cs:-100,1) -- (axis cs:1000,1);

            %% Label
            \node[font=\footnotesize,anchor=south west] () at (axis cs:5,0) {(d) Sensor 4};

            %% Legend
            \legend{RBF-50, RQ-200, Matérn-200}

    \end{groupplot}
\end{tikzpicture}
    \caption{Online drift correction: median relative MSE for each sensor at different calibration intervals.}
    \label{fig:drift_correction_online}
\end{figure*}
%%%%%%%%%%%%%%%%%%%%%%%%%%%%%%%%%%%%%%%%%%%%%%%%%%%

For online drift correction, the GPR models are trained only on previous calibrations.
Thus, each time a calibration is performed, the GPR models are re-trained and used for drift correction until the next calibration.
Results for online drift correction are shown in \cref{fig:drift_correction_online} in terms of relative MSE compared to stepwise interpolation.
Note that linear interpolation is not relevant for online drift correction since it requires knowledge of the next calibration.

The increased challenge of online drift correction compared to offline is clearly evident in our results, which show no significant difference between our proposed methods and stepwise interpolation for three out of the four sensors.
That is, for sensors 2, 3 and 4, our best performing kernel, Matérn, delivers average relative MSE scores of 0.98, 1.01 and 0.98, respectively, indicating that online drift correction is ineffective for these sensors.
% and stepwise interpolation may be the best online approach.

Nonetheless, we are able to significantly reduce online MSE for sensor 1 by 30\% on average across the range of calibration invervals tested using the Matérn kernel.
However, performance is highly dependent on the calibration interval, with greater improvements possible for shorter intervals, as shown in \cref{fig:calibration_timing_optimisation_aggregate}(a).

\subsection{Calibration Schedule Optimisation}

%%%%%%%%%%%%%%%%%%%%%%%%%%%%%%%%%%%%%%%%%%%%%%%%%%%
% Multi-sensor calibration timing optimisation results: all sensors
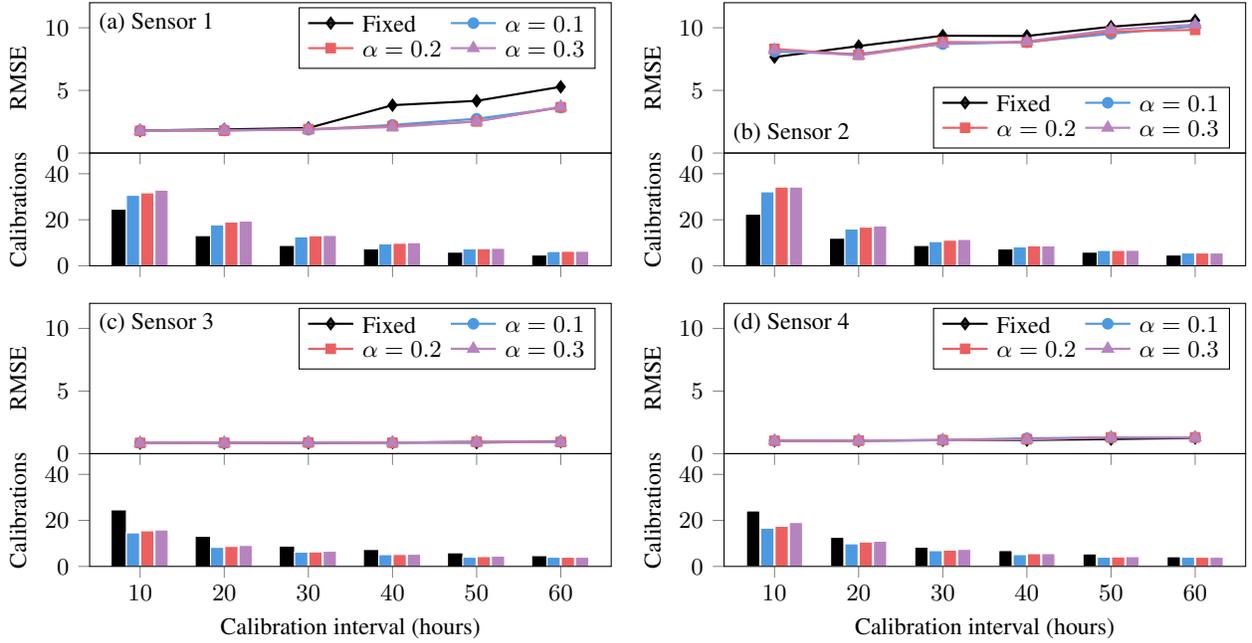
\begin{figure*}
    \centering
    \newcommand{\xmin}{4}

\begin{tikzpicture}
    \begin{groupplot}[
        group style={group size=2 by 4,xlabels at=edge bottom,vertical sep=0.5cm,horizontal sep=1.5cm,xticklabels at=edge bottom},
        width=0.42\columnwidth,
        height=2.0cm,
        xmin=\xmin,xmax=66,
        ymin=0,ymax=12,
        xtick={10, 20, 30, 40, 50, 60},
        xlabel={Calibration interval (hours)},
        ylabel={RMSE},
        legend columns=2,
        legend pos=north east,
        ]
        %%%%%%%%%%%%%%%%%%%%%%%%%%%%%%%%%%%%%%%%%%%%%%%%%%%%%%
        % Sensor 1: MSE
        \nextgroupplot[generic line plot]
            \addplot [black, thick, mark=diamond*] table[x index=0, y index=1] {data__calibration_schedule_optimisation_by_sensor_rmse.csv};
            \pgfplotsset{cycle list shift=-1}
            \addplot+ [] table[x index=0, y index=2] {data__calibration_schedule_optimisation_by_sensor_rmse.csv};
            \addplot+ [] table[x index=0, y index=3] {data__calibration_schedule_optimisation_by_sensor_rmse.csv};
            \addplot+ [] table[x index=0, y index=4] {data__calibration_schedule_optimisation_by_sensor_rmse.csv};

            %% Label
            \node[anchor=north west] () at (axis cs:\xmin,12) {(a) Sensor 1};

            %% Legend
            \legend{Fixed, $ \alpha = 0.1 $, $ \alpha = 0.2 $, $ \alpha = 0.3 $}

        %%%%%%%%%%%%%%%%%%%%%%%%%%%%%%%%%%%%%%%%%%%%%%%%%%%%%%
        % Sensor 2: MSE
        \nextgroupplot[generic line plot,legend pos=south east]
            \addplot [black, thick, mark=diamond*] table[x index=0, y index=5] {data__calibration_schedule_optimisation_by_sensor_rmse.csv};
            \pgfplotsset{cycle list shift=-1}
            \addplot+ [] table[x index=0, y index=6] {data__calibration_schedule_optimisation_by_sensor_rmse.csv};
            \addplot+ [] table[x index=0, y index=7] {data__calibration_schedule_optimisation_by_sensor_rmse.csv};
            \addplot+ [] table[x index=0, y index=8] {data__calibration_schedule_optimisation_by_sensor_rmse.csv};

            %% Label
            \node[anchor=south west] () at (axis cs:\xmin,0) {(b) Sensor 2};

            %% Legend
            \legend{Fixed, $ \alpha = 0.1 $, $ \alpha = 0.2 $, $ \alpha = 0.3 $}

        %%%%%%%%%%%%%%%%%%%%%%%%%%%%%%%%%%%%%%%%%%%%%%%%%%%%%%
        % Sensor 1: no. calibrations
        \nextgroupplot[generic bar plot,ymax=49,height=1.5cm,ylabel=Calibrations,yshift=0.5cm,enlarge x limits=0,bar width=5pt]
            \addplot [fill=black] table[x index=0, y index=1] {data__calibration_schedule_optimisation_by_sensor_n_calibrations.csv};
            \pgfplotsset{cycle list shift=-1}
            \addplot+ [] table[x index=0, y index=2] {data__calibration_schedule_optimisation_by_sensor_n_calibrations.csv};
            \addplot+ [] table[x index=0, y index=3] {data__calibration_schedule_optimisation_by_sensor_n_calibrations.csv};
            \addplot+ [] table[x index=0, y index=4] {data__calibration_schedule_optimisation_by_sensor_n_calibrations.csv};

        %%%%%%%%%%%%%%%%%%%%%%%%%%%%%%%%%%%%%%%%%%%%%%%%%%%%%%
        % Sensor 2: no. calibrations
        \nextgroupplot[generic bar plot,ymax=49,height=1.5cm,ylabel=Calibrations,yshift=0.5cm,enlarge x limits=0,bar width=5pt]
            \addplot [fill=black] table[x index=0, y index=5] {data__calibration_schedule_optimisation_by_sensor_n_calibrations.csv};
            \pgfplotsset{cycle list shift=-1}
            \addplot+ [] table[x index=0, y index=6] {data__calibration_schedule_optimisation_by_sensor_n_calibrations.csv};
            \addplot+ [] table[x index=0, y index=7] {data__calibration_schedule_optimisation_by_sensor_n_calibrations.csv};
            \addplot+ [] table[x index=0, y index=8] {data__calibration_schedule_optimisation_by_sensor_n_calibrations.csv};

        %%%%%%%%%%%%%%%%%%%%%%%%%%%%%%%%%%%%%%%%%%%%%%%%%%%%%%
        % Sensor 3: MSE
        \nextgroupplot[generic line plot]
            \addplot [black, thick, mark=diamond*] table[x index=0, y index=9] {data__calibration_schedule_optimisation_by_sensor_rmse.csv};
            \pgfplotsset{cycle list shift=-1}
            \addplot+ [] table[x index=0, y index=10] {data__calibration_schedule_optimisation_by_sensor_rmse.csv};
            \addplot+ [] table[x index=0, y index=11] {data__calibration_schedule_optimisation_by_sensor_rmse.csv};
            \addplot+ [] table[x index=0, y index=12] {data__calibration_schedule_optimisation_by_sensor_rmse.csv};

            %% Label
            \node[anchor=north west] () at (axis cs:\xmin,12) {(c) Sensor 3};

            %% Legend
            \legend{Fixed, $ \alpha = 0.1 $, $ \alpha = 0.2 $, $ \alpha = 0.3 $}

        %%%%%%%%%%%%%%%%%%%%%%%%%%%%%%%%%%%%%%%%%%%%%%%%%%%%%%
        % Sensor 4: MSE
        \nextgroupplot[generic line plot]
            \addplot [black, thick, mark=diamond*] table[x index=0, y index=13] {data__calibration_schedule_optimisation_by_sensor_rmse.csv};
            \pgfplotsset{cycle list shift=-1}
            \addplot+ [] table[x index=0, y index=14] {data__calibration_schedule_optimisation_by_sensor_rmse.csv};
            \addplot+ [] table[x index=0, y index=15] {data__calibration_schedule_optimisation_by_sensor_rmse.csv};
            \addplot+ [] table[x index=0, y index=16] {data__calibration_schedule_optimisation_by_sensor_rmse.csv};

            %% Label
            \node[anchor=north west] () at (axis cs:\xmin,12) {(d) Sensor 4};

            %% Legend
            \legend{Fixed, $ \alpha = 0.1 $, $ \alpha = 0.2 $, $ \alpha = 0.3 $}

        %%%%%%%%%%%%%%%%%%%%%%%%%%%%%%%%%%%%%%%%%%%%%%%%%%%%%%
        % Sensor 3: no. calibrations
        \nextgroupplot[generic bar plot,ymax=49,height=1.5cm,ylabel=Calibrations,yshift=0.5cm,enlarge x limits=0,bar width=5pt]
            \addplot [fill=black] table[x index=0, y index=9] {data__calibration_schedule_optimisation_by_sensor_n_calibrations.csv};
            \pgfplotsset{cycle list shift=-1}
            \addplot+ [] table[x index=0, y index=10] {data__calibration_schedule_optimisation_by_sensor_n_calibrations.csv};
            \addplot+ [] table[x index=0, y index=11] {data__calibration_schedule_optimisation_by_sensor_n_calibrations.csv};
            \addplot+ [] table[x index=0, y index=12] {data__calibration_schedule_optimisation_by_sensor_n_calibrations.csv};

        %%%%%%%%%%%%%%%%%%%%%%%%%%%%%%%%%%%%%%%%%%%%%%%%%%%%%%
        % Sensor 4: no. calibrations
        \nextgroupplot[generic bar plot,ymax=49,height=1.5cm,ylabel=Calibrations,yshift=0.5cm,enlarge x limits=0,bar width=5pt]
            \addplot+ [fill=black] table[x index=0, y index=13] {data__calibration_schedule_optimisation_by_sensor_n_calibrations.csv};
            \pgfplotsset{cycle list shift=-1}
            \addplot+ [] table[x index=0, y index=14] {data__calibration_schedule_optimisation_by_sensor_n_calibrations.csv};
            \addplot+ [] table[x index=0, y index=15] {data__calibration_schedule_optimisation_by_sensor_n_calibrations.csv};
            \addplot+ [] table[x index=0, y index=16] {data__calibration_schedule_optimisation_by_sensor_n_calibrations.csv};

    \end{groupplot}
\end{tikzpicture}
    \caption{
        Average MSE and number of calibrations for each sensor comparing fixed-interval calibration scheduling with our proposed adaptive method.
    }
    \label{fig:calibration_timing_optimisation_all_sensors}
\end{figure*}
%%%%%%%%%%%%%%%%%%%%%%%%%%%%%%%%%%%%%%%%%%%%%%%%%%%

%%%%%%%%%%%%%%%%%%%%%%%%%%%%%%%%%%%%%%%%%%%%%%%%%%%
% Multi-sensor calibration schedule optimisation - aggregate - figure & table
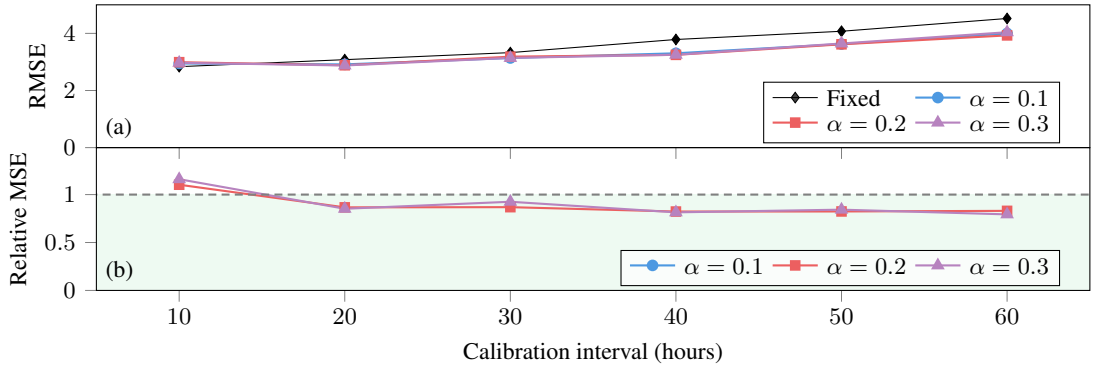
\begin{figure}
    \centering
    \begin{tikzpicture}
    \begin{groupplot}[
        group style={group size=1 by 2,xlabels at=edge bottom,vertical sep=0cm,xticklabels at=edge bottom},
        generic line plot,
        width=0.8\columnwidth,
        height=1.9cm,
        ymin=0,
        xmin=5,xmax=65,
        xlabel={Calibration interval (hours)},
        legend pos=south east,
        legend columns=2,
        ]
        % Average MSE
        \nextgroupplot[ymax=5,ylabel={RMSE}]
            \addplot [black, mark=diamond*] table[x index=0, y index=1] {data__calibration_schedule_optimisation_aggregate.csv};
            \pgfplotsset{cycle list shift=-1}
            \addplot+ [] table[x index=0, y index=2] {data__calibration_schedule_optimisation_aggregate.csv};
            \addplot+ [] table[x index=0, y index=3] {data__calibration_schedule_optimisation_aggregate.csv};
            \addplot+ [] table[x index=0, y index=4] {data__calibration_schedule_optimisation_aggregate.csv};

            % Label
            \node[anchor=south west] () at (axis cs: 5,0) {(a)};

            % Legend
            \legend{Fixed, $ \alpha = 0.1 $, $ \alpha = 0.2 $, $ \alpha = 0.3 $};

        % Normalised MSE
        \nextgroupplot[ymax=1.49,ylabel={Relative MSE},ytick={0,0.5,1},legend columns=3]
            % Fill good region
            \draw[colour4!10, fill] (axis cs:5,0) -- (axis cs:65,0) -- (axis cs:65,1) -- (axis cs:5,1) -- cycle;

            % Plot data
            \addplot+ [] table[x index=0, y index=4] {data__calibration_schedule_optimisation_aggregate.csv};
            \addplot+ [] table[x index=0, y index=5] {data__calibration_schedule_optimisation_aggregate.csv};
            \addplot+ [] table[x index=0, y index=6] {data__calibration_schedule_optimisation_aggregate.csv};

            % Annotate reference line
            \draw[densely dashed, thick, black!50] (axis cs:-100,1) -- (axis cs:1000,1);

            % Label
            \node[anchor=south west] () at (axis cs: 5,0) {(b)};

            % Legend
            \legend{$ \alpha = 0.1 $, $ \alpha = 0.2 $, $ \alpha = 0.3 $}
    \end{groupplot}
\end{tikzpicture}
    \captionof{figure}{
        Average RMSE and relative MSE compared to fixed-interval calibration scheduling for selected values of $ \learningrate $ over all sensors.}
    \label{fig:calibration_timing_optimisation_aggregate}
\end{figure}
\begin{table}
    \centering
    \caption{
        Average RMSE and relative MSE compared to fixed-interval calibration scheduling for selected values of $ \learningrate $ over all sensors.
        Best performance highlighted in \textbf{bold}.
    }
    \footnotesize

\begin{tabular}{r|rrrr|rrr}
    \toprule
             &              \multicolumn{4}{c|}{\textbf{RMSE}}               &   \multicolumn{3}{c}{\textbf{Relative MSE}}   \\
    Interval &         Fixed &           0.1 &           0.2 &           0.3 &           0.1 &           0.2 &           0.3 \\ \midrule
          10 & \textbf{2.83} &          2.95 &          2.99 &          2.97 & \textbf{1.10} &          1.16 &          1.14 \\
          20 &          3.08 &          2.91 &          2.88 & \textbf{2.88} &          0.87 &          0.85 & \textbf{0.85} \\
          30 &          3.33 & \textbf{3.14} &          3.18 &          3.14 & \textbf{0.87} &          0.93 &          0.89 \\
          40 &          3.79 &          3.30 & \textbf{3.25} &          3.25 &          0.82 & \textbf{0.81} &          0.83 \\
          50 &          4.07 &          3.63 & \textbf{3.62} &          3.65 & \textbf{0.82} &          0.84 &          0.87 \\
          60 &          4.52 &          4.00 & \textbf{3.93} &          4.04 &          0.83 & \textbf{0.79} &          0.85 \\ \midrule
         All &          3.60 &          3.32 & \textbf{3.31} &          3.32 & \textbf{0.89} &          0.90 &          0.90 \\ \bottomrule
\end{tabular}

    \label{tab:calibration_timing_optimisation_aggregate}

\end{table}
%%%%%%%%%%%%%%%%%%%%%%%%%%%%%%%%%%%%%%%%%%%%%%%%%%%

To evaluate calibration scheduling, each sensor is initially assigned the same calibration interval, which dictates the overall calibration budget.
%For example, if the initial calibration interval is 20~hours, then the budget is one calibration per 20 hours per sensor, or one every five hours across all four sensors.
Calibration intervals are then updated periodically for each sensor according to Algorithm 1 and \cref{sec:method:calibration_interval_optimisation}.
An update frequency of once per hour was used.

We tested calibration intervals from 10 to 60 in increments of 10 hours and alpha values from 0.05 to 0.4 in increments of 0.05.
Each combination was repeated 100 times with random initial calibration times, as is \cref{sec:evaluation:driftcorrection}.
Mean results for each sensor are shown in \cref{fig:calibration_timing_optimisation_all_sensors} for a selection of $ \learningrate $ value.
The upper part of each sub-figure shows RMSE and the lower part shows the average number of calibrations.
The black data series corresponds to using a fixed calibration interval ($ \learningrate = 0 $).

As shown in \cref{fig:calibration_timing_optimisation_all_sensors}, unstable sensors (i.e., sensors 1 and 2) are allocated more calibrations and, as a result, achieve lower RMSE values than when using a fixed-interval calibration schedule.
Conversely, stable, low-drift sensors are allocated fewer calibrations and achieve either the same or slightly higher RMSE values.
However, critically, the reduction in RMSE for high-drift sensors outweighs the increase on the low-drift sensors.
This can be seen by comparing \cref{fig:calibration_timing_optimisation_all_sensors}(a), which shows a significant RMSE reduction of up to 1.68 or 44\% for sensor 1 (at a 40 hour initial interval), with \cref{fig:calibration_timing_optimisation_all_sensors}(c) and (d), which show negligible increases in RMSE of no more than 0.14.

The net effect across all sensors is shown in \cref{fig:calibration_timing_optimisation_aggregate} and \cref{tab:calibration_timing_optimisation_aggregate}.
As shown in \cref{fig:calibration_timing_optimisation_aggregate}(a), apart from the 10~hour initial interval case, where overall RMSE is high due to Sensor 2, our approach successfully reduces overall RMSE across all calibration budgets compared to a fixed-interval calibration schedule.
\cref{fig:calibration_timing_optimisation_aggregate}(b) and the right side of \cref{tab:calibration_timing_optimisation_aggregate} show the same data expressed as relative MSE compared to the fixed interval case.
Although there is little difference between different values of $ \learningrate $, it appears that $ \learningrate = 0.1 $ performs best with a mean reduction in overall MSE of 11.4\%.
Excluding 10~hour initial intervals, this increases to 15.7\%.

    \section{Conclusion}
\label{sec:conclusion}

In this paper, we have proposed a novel method for sensor drift correction in chemical sensors using GPR to directly model the individual coefficients in the sensor response function.
We have shown that this approach can significantly and consistently reduce MSE in offline settings through our tests on DO sensors.
Despite presenting a harder challenge, online drift correction can also be done using our approach, where it performs at least as well as baseline techniques.
We also proposed a unique uncertainty-driven calibration scheduling method that dynamically adjusts calibration frequency at the sensor level to optimise network-wide accuracy.
Our test results on DO sensors show that this can further reduce MSE by allocating more calibrations to high-drift and unstable sensors.

    \section{Acknowledgements}

This work was supported in part by a research grant (VIL 50075) from Villum Fonden and an AIAS-AUFF Fellowship (AK) from the Aarhus Institute of Advanced Studies and Aarhus Universitets Forskningsfond.
We would also like to thank the Grundfos Foundation, as well as Lars Borregaard Pedersen, who helped with the measurements.

    \bibliographystyle{ieeetran}
    \bibliography{library}

\end{document}